\documentclass[journal]{IEEEtran}
\usepackage[utf8]{inputenc}
\usepackage{bm}
\usepackage{amsmath,epsfig,amssymb,dsfont,amsthm}
\usepackage{hyperref}
\usepackage{verbatim}
\usepackage{caption}
\usepackage{subcaption}
\usepackage[dvipsnames]{xcolor}
\usepackage{graphicx}
	\graphicspath{{./figures/}}
\usepackage{float}
\usepackage{url}
\usepackage{eurosym}
\usepackage{setspace}
\usepackage{balance}
\usepackage{dsfont}
\usepackage{bbm}
\usepackage{array}
\usepackage{cite}
\usepackage[ruled,vlined]{algorithm2e}
\usepackage{thmtools}
\usepackage{comment}
\usepackage[normalem]{ulem}

\usepackage{tabularx,booktabs}
\newcolumntype{A}{<{\raggedright\arraybackslash}X}
\newcolumntype{B}{>{\raggedleft\arraybackslash}{\hsize=.3\hsize}X}
\newcolumntype{C}{>{\hsize=.24\hsize}X}

\newcommand{\blambda}{\boldsymbol{\Lambda}}
\newcommand{\bA}{\boldsymbol{A}}
\newcommand{\bL}{\boldsymbol{\mathcal L}}
\newcommand{\cL}{\mathcal L}
\newcommand{\R}{\mathcal R}
\newcommand{\bT}{\mathsf{T}}
\newcommand{\bE}{\mathbb{E}}
\newcommand{\bF}{\boldsymbol{\mathcal F}}
\newcommand{\bmu}{\boldsymbol \mu}
\newcommand{\bpsi}{\boldsymbol \psi}
\newcommand{\bzeta}{\boldsymbol \zeta}
\newcommand{\bdelta}{\boldsymbol \Delta}
\newcommand{\Tr}{\textrm{Tr}}

\theoremstyle{plain}
\newtheorem{Thm}{Theorem}

\newtheorem{Lem}{Lemma}
\newtheorem{Asm}{Assumption}

\theoremstyle{remark}

\declaretheoremstyle[%
 spaceabove=-6pt,%
 spacebelow=6pt,%
 headfont=\normalfont\itshape,%
 postheadspace=1em,%
 qed=\qedsymbol%
]{mystyle} 
\declaretheorem[name={Proof},style=mystyle,unnumbered,
]{prf}

\newcommand{\qedsymb}{\hfill\ensuremath{\blacksquare}}                 
\renewcommand{\qedsymbol}{\qedsymb}                                    
\allowdisplaybreaks
\title{Discovering Influencers in Opinion Formation over Social Graphs} 
\author{\IEEEauthorblockN{Valentina Shumovskaia, Mert Kayaalp, Mert Cemri, and Ali H. Sayed}
\thanks{V. Shumovskaia, M. Kayaalp, and A. H. Sayed are with the École Polytechnique Fédérale de Lausanne (EPFL). Emails: $\{$valentina.shumovskaia, mert.kayaalp, ali.sayed$\}$@epfl.ch. M. Cemri is with the Department of Electrical and Electronics Engineering, Bilkent University, 06800, Ankara, Turkey. The author contributed to the work as a visiting student at EPFL, Switzerland. E-mail: mert.cemri@ug.bilkent.edu.tr.

This work was supported in part by grant 205121-184999 from the Swiss National Science Foundation. A short version of this work appears in the conference publication~\cite{shumovskaia2023}.
}
}

\IEEEoverridecommandlockouts
\begin{document}

\maketitle

\begin{abstract}
    The adaptive social learning paradigm helps model how networked agents are able to form opinions on a state of nature and track its drifts in a changing environment. In this framework, the agents repeatedly update their beliefs based on private observations and exchange the beliefs with their neighbors. In this work, it is shown how the sequence of publicly exchanged beliefs over time allows users to discover rich information about the underlying network topology and about the flow of information over the graph. In particular, it is shown that it is possible (i) to identify the influence of each individual agent to the objective of truth learning, (ii) to discover how well-informed each agent is, (iii) to quantify the pairwise influences between agents, and (iv) to learn the  underlying network topology. The algorithm derived herein is also able to work under non-stationary environments where either the true state of nature or the graph topology are allowed to drift over time. We apply the proposed algorithm to different subnetworks of Twitter users, and identify the most influential and central agents by using their public tweets (posts).
\end{abstract}

\begin{IEEEkeywords}
Social learning, social influence, explainability, inverse modeling, online learning, graph learning, Twitter.
\end{IEEEkeywords}


\maketitle

\section{Introduction And Related Work}
    The social learning paradigm is a popular non-Bayesian formulation that enables a group of  networked agents to learn and track the state of nature. It has motivated several studies in the literature with many useful variations under varied modeling assumptions  (see, e.g.,~\cite{jadbabaie2012non, nedic2017fast, molavi2017foundations, molavi2018theory, bordignon2020adaptive, bordignon2020social, lalitha2018social, zhao2012learning, gale2003bayesian, acemoglu2011bayesian, inan2022social, toghani2021communication, 9132712, 9670665, 9670668, kayaalp2022distributed}). Under this framework, agents observe streaming data and share information with their immediate neighbors. Through a process of localized cooperation, the agents continually update their {\em beliefs} about the underlying state. These beliefs describe the agents' confidence on each possible hypothesis. 
    The main question in social learning is whether agents are able to learn the truth eventually, i.e., whether the beliefs on the wrong hypotheses vanish.
    
    In principle, each agent in the network could consider pursuing a {\em fully} Bayesian solution to learn and track the state of nature. However, this solution is intractable and generally NP-hard~\cite{gale2003bayesian, acemoglu2011bayesian, hkazla2021bayesian}. This is because it requires that each agent has access to the data from  the entire network, in addition to their knowledge of the full graph topology. These pieces of information are rarely available in a decentralized setting. For this reason, non-Bayesian approaches have been devised as effective alternatives~\cite{jadbabaie2012non, nedic2017fast, molavi2017foundations, molavi2018theory, bordignon2020adaptive, bordignon2020social, lalitha2018social, zhao2012learning}. 
    In this formulation, the agents first perform a local Bayesian update using their newly received private observations, and then fuse their beliefs with those of their neighbors either linearly or geometrically~\cite{nedic2017fast,bordignon2020adaptive,lalitha2018social,zhao2012learning,kayaalp2022arithmetic}.
    This approach allows for diverse data models across the agents, and helps preserve the privacy of the  individual observations. 
    In this work, we adopt the {\em Adaptive Social Learning} (ASL) strategy from~\cite{bordignon2020adaptive}, which showed how to extend traditional non-Bayesian learning under {\em fixed} truth to {\em dynamic} scenarios where the state of nature is allowed to drift with time. Under ASL, the agents will be able to track these drifts rather effectively with performance guarantees.

    Now, given a collection of networked agents tracking the state of nature by means of the adaptive social learning (ASL) strategy, our main objective is to focus on two questions related to {\em explainability} and {\em inverse modeling}. In particular, by observing the sequence of publicly exchanged beliefs, we would like to discover the underlying graph topology (i.e., how the agents are connected to each other). 
    We would also like to discover each agent's contribution (or influence) to the network's learning process.
    
    The question of explainability over graphs is actually a challenging task, and it has been receiving  increasing attention (e.g.,~\cite{rudin2022interpretable, heuillet2022collective, ohana2021explainable, juozapaitis2019explainable, 9355456, brandao2022explainability, georgara2022building}).  
    These works approach explainability from different perspectives, and with different aims. 
    For instance, in~\cite{brandao2022explainability, georgara2022building}, the authors aim at building a framework for human understanding of why a black-box method arrives at a particular solution. 
    The work~\cite{rudin2022interpretable} argues that a better understanding of multi-agent reinforcement learning can help to limit the search space. 
    Other works suggest modifying the learning algorithm for better interpretability~\cite{juozapaitis2019explainable}. 
    Overall, higher transparency and a better  understanding of the solutions are generally crucial for critical applications, such as using artificial intelligence in healthcare or in autonomous devices (such as vehicles).
    
    We have performed some prior work on explainability for social learning~\cite{shumovskaia2022explainability, shumovskaia}. 
    In these works, given the evolution of beliefs and assuming some partial prior knowledge about the distribution of private observations, it was shown  how to identify pairwise influences inside the network (i.e., how strongly pairs of agents influence each other), as well as how to recover the underlying graph topology. 
    One of the key differences with the previous works is that here, we consider a more {\em limited} (and, therefore, more challenging) information scenario, where we do not require any information about the probability distribution of the observations (or their likelihoods).
    Despite being more challenging to analyse, the algorithm nevertheless becomes more practical and applicable to real-world data, where distributions of privately received observations remain unknown.
    We will show that, under this more demanding scenario, we are still able to identify the contribution of each agent to truth learning, assess its level of informativeness, as well as learn the underlying graph and the pairwise influences between agents.  
    The modeling conditions we consider are generally commonplace in real-world social networks, such as Twitter~\cite{polisci_twitter, twitter_quercia2011mood, twitter_influence_measure, twitter_influence_swedish, pennacchiotti2011machine, lin2012large, hasan2018machine, zervoudakis2021opinionmine}. We consider a Twitter application in Section~\ref{sec:twitter}. 
    In this application, users publicly exchange their opinions on Twitter, which are therefore observable. However, we do not have access to additional information about sources affecting people's opinion during these exchanges such as News or discussions occurring outside Twitter and, therefore, we do not have information about the signal distributions. 
    Identifying the most influential users and their communication patterns can provide valuable insights in social network analysis~\cite{camacho2020four}.
    Actually, the problem of identifying the most influential nodes in a network is increasingly relevant~\cite{kempe2003maximizing, kempe2005influential, li2018influence, gomez2016influence, qiu2018deepinf, karoui2022machine}, especially following the rise of online social networks. Once identified, this information can be useful in many contexts. For example, it can be used to enhance recommendations for marketing purposes~\cite{kempe2003maximizing, kempe2005influential, li2018influence, gomez2016influence, qiu2018deepinf, karoui2022machine,domingos2001mining, saito2008prediction, granovetter1978threshold, chen2012time, wu2019neural, goldenberg2001talk}, where the objective is to maximize the number of influenced nodes. 
    
    There have also been several important contributions to graph learning such as~\cite{shumovskaia, shumovskaia2022explainability, kalofolias2016learn, egilmez2017graph, vlaski2018online, dong2019learning, pasdeloup2017characterization, thanou2017learning, chepuri2017learning, shafipour2017network, segarra2016network, sardellitti2016graph, maretic2017graph, cirillo2021learning, yiran, saboksayr2021accelerated}. 
    This is because the graph structure plays an important role in distributed learning~\cite{8737602, shahrampour2015distributed, hu2022optimal, 9130856}, and its identification can provide valuable information about relationships within the network~\cite{shumovskaia2022explainability}.
    One approach is based on exploiting correlations and similarities between vertices~\cite{friedman2008sparse, matta2019graph, brovelli2004beta}.
    Other works assume a defined model behind the graph signal, such as the heat diffusion process~\cite{pasdeloup2017characterization, vlaski2018online, thanou2017learning}.
    Structural constraints (such as sparsity, connectivity, symmetry) can also be introduced through regularization~\cite{egilmez2017graph, chepuri2017learning, maretic2017graph}. 
    The present work examines the problem of identifying a graph in the adaptive social learning setting.
    Learning the graph from social interactions requires a different approach, as already illustrated in~\cite{shumovskaia}, due to the special form of the non-stationary observation signals. Importantly, our study will relax certain assumptions introduced in that work.
    
    The paper is organized as follows. We describe the adaptive social learning model in Section~\ref{sec:model}. 
    Then, we propose an algorithm for learning the combination matrix and the agents' informativeness in Section~\ref{sec:algorithm}.
    In Section~\ref{sec:influence_identification}, we justify the fact that each agent's contribution to truth learning is proportional to its relative centrality and its informativeness (the KL-divergence between the marginal likelihood of the truth and the other hypotheses).
    We provide theoretical performance guarantees of the algorithm in Section~\ref{sec:theory}.
    In Section~\ref{sec:experiments}, we illustrate  performance in different settings by means of numerical simulations. Finally, in Section~\ref{sec:twitter}, we apply the algorithm to real-world data from Twitter.
    
\section{Social Learning Model}\label{sec:model}
\begin{figure}
        \centering
        \includegraphics[width=0.75\linewidth]{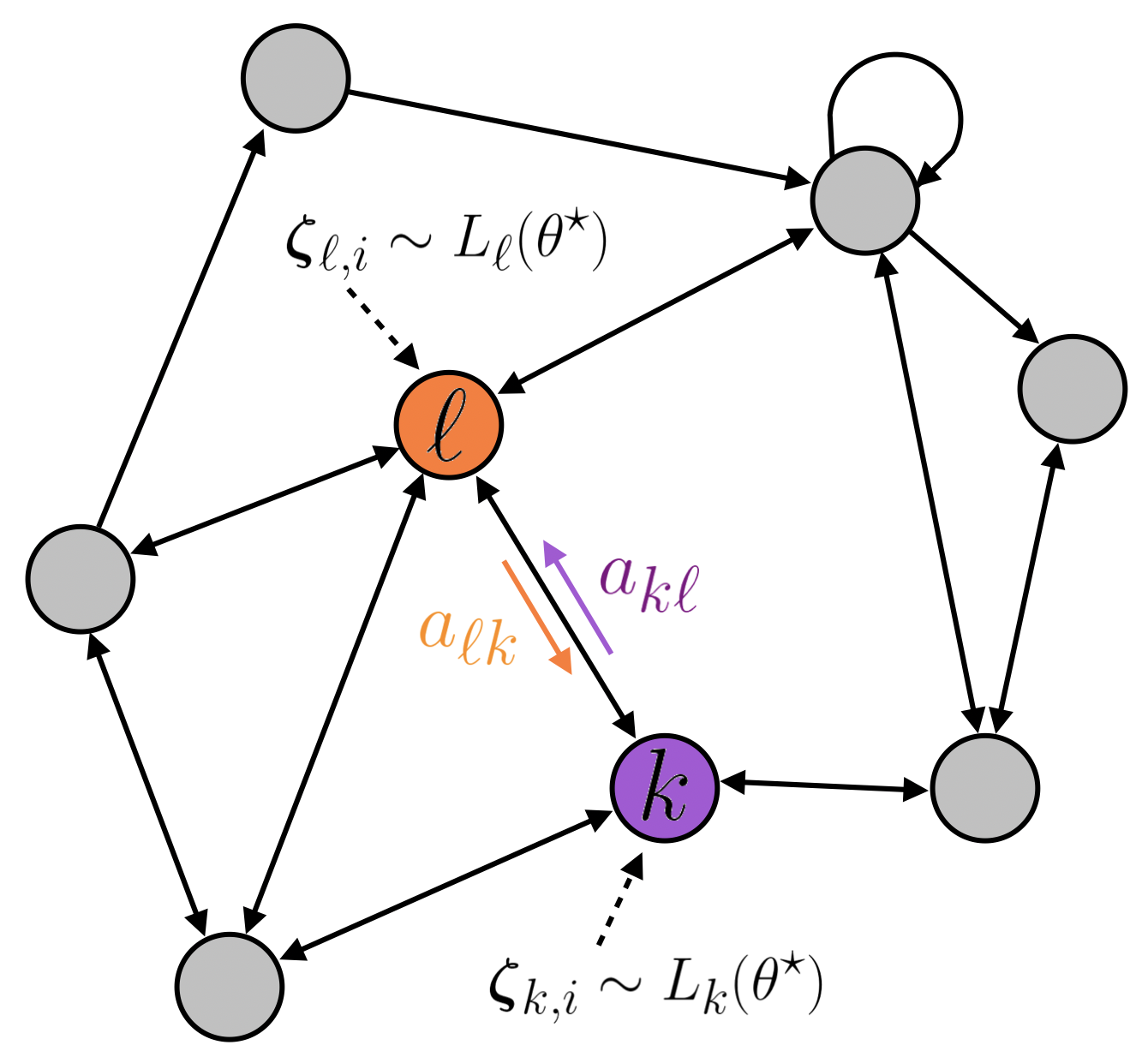}
        \caption{An illustration of the network model.}
        \label{fig:scheme}
    \end{figure}

    We refer to Fig.~\ref{fig:scheme} and consider a collection of agents $\mathcal N$ performing peer-to-peer exchanges of beliefs according to some combination matrix $A_\star$ with non-negative entries, $[A_\star]_{\ell,k} = a_{\ell k} \geq 0$. 
    Agent $\ell$ is able to communicate with agent $k$ when $a_{\ell k}$ is positive; this scalar refers to the weight that agent $k$ assigns to the information received from agent $\ell$. 
    We assume the matrix $A_{\star}$ is  \textit{left-stochastic} and corresponds to a \textit{strongly connected} graph~\cite{lalitha2018social, nedic2017fast, bordignon2020adaptive}. 
    The first assumption means that the entries on any  matrix column $k \in \mathcal N$ add up to one, $\sum_{\ell \in \mathcal N}a_{\ell k} = 1$. 
    The second assumption means that there exists a path with positive weights between any two agents, and there is at least one agent in the network that does not ignore its own observation, i.e., $a_{kk}>0$ for at least one $k\in {\mathcal N}$.
    This implies that the combination matrix is primitive, i.e., for any $\ell$, $k \in \mathcal N$, there exists $t > 0$ such that $[A_\star^t]_{\ell, k} > 0$.
    It follows from the Perron-Frobenius theorem ~\cite[Chapter 8]{horn2009},~\cite{sayed_2023} that the power matrix $A_{\star}^t$ converges to $u\mathds 1^\bT$ as $t \rightarrow \infty$ at an exponential rate, where $u$ is the Perron eigenvector that satisfies:
    \begin{align}
        A_\star u = u,\qquad u_\ell > 0,\qquad \sum_{\ell \in \mathcal N} u_\ell = 1,
    \end{align}
    where the $u_{\ell}$ denote the individual entries of $u$. Each of these entries describes the centrality of the corresponding agent in the graph (i.e., its level of contribution to the inference task). 
    
    We assume that there exists one true state of nature $\theta^\star$ belonging to a finite set of hypotheses, denoted by $\Theta$. 
    Initially, each agent $k$ starts with a \textit{private belief} vector $\bmu_{k,0} \in [0,1]^{|\Theta|}$, where each entry $\bmu_{k,0}(\theta)$ describes how confident agent $k$ is that $\theta$ corresponds to the true hypothesis $\theta^{\star}$.
    As befitting of a true probability mass function, the total confidence sums up to one, $\sum_{\theta}\bmu_{k,0}(\theta) = 1$. 
    To ensure that no hypothesis is excluded beforehand by any of the agents, we assume that $\bmu_{k,0}(\theta) > 0$, $\forall \theta \in \Theta$.
    
    At each time instant $i$, each agent $k$ observes a measurement $\bzeta_{k,i}$. We assume initially that each agent $k\in\mathcal N$ has access to private  likelihood functions, $L_{k}(\zeta|\theta)$, which describe the distribution of the observation $\zeta$ conditioned on each potential model $\theta$. With a slight abuse of notation, we sometimes denote $L_k(\cdot|\theta)$ by $L_k(\theta)$.
    The observations $\bzeta_{k, i}$ are assumed to be independent and identically distributed (i.i.d.) over time. 
    In order to be able to distinguish the true hypothesis $\theta^\star$ from any other hypothesis $\theta \neq \theta^\star$, we need to assume that for any $\theta\neq \theta^{\star}$, there exists at least one clear-sighted agent $k\in \mathcal N$ that has strictly positive KL-divergence relative to the true likelihood, i.e., $D_{\textrm{KL}}\left(L_k\left(\theta^\star\right)||L_k\left(\theta\right)\right) > 0$.
    The following boundedness assumption on the likelihood is common in the literature~\cite{bordignon2020social, shumovskaia2022explainability}; it essentially amounts to assuming that the likelihoods share support regions. 
    \begin{Asm}[\bf{Bounded likelihoods}]
        \label{asm:support}
        There exists a finite constant $b > 0$ such that, for all $k \in \mathcal N$:
        \begin{align}
            \Bigg|\log \frac {L_k(\boldsymbol\zeta | \theta)}{L_k(\boldsymbol\zeta | \theta')} \Bigg| \leq b
        \end{align}
        for all $\theta,\;\theta' \in \Theta$, and $\boldsymbol\zeta$.
        \newline $\textrm{ }$
        \qedsymb
    \end{Asm}
    
    Next, we describe the ASL strategy from~\cite{bordignon2020adaptive}.
    At each time step $i$, each agent $k$ performs a local update based on the newly received observation and forms the \textit{intermediate (public)} belief:
    \begin{align}
        &\boldsymbol{\psi}_{k,i}(\theta)=
        \frac{L_k^\delta(\boldsymbol{\zeta}_{k,i}|\theta)\boldsymbol{\mu}^{1-\delta}_{k,i-1}(\theta)}{\sum_{\theta'\in\Theta}L_k^{\delta}(\boldsymbol{\zeta}_{k,i}|\theta')\boldsymbol{\mu}^{1-\delta}_{k,i-1}(\theta')},\quad k\in\mathcal{N}.\label{eq:adapt_adaptive}
    \end{align}
    \noindent 
    Here, $\delta \in (0,1) $ is a step-size parameter that controls the adaptation capacity, i.e., $\delta$ controls the importance of newly received data relative to the past history. This intermediate belief is shared with the \emph{follower} agents of $k$, i.e., with all agents $\ell$ for which $a_{k\ell}>0$. Subsequently, agent $k$ fuses the beliefs received from its neighbors, i.e., from all agents for which $a_{\ell k}>0$. We denote this set by ${\mathcal N}_k$. One fusion rule is to fuse the beliefs geometrically as follows in order to obtain the {\em private} beliefs~\cite{nedic2017fast, bordignon2020adaptive, lalitha2018social, kayaalp2022arithmetic}:
    \begin{align}
        &\boldsymbol{\mu}_{k,i}(\theta)=\frac{\prod_{\ell\in\mathcal{N}_k}\boldsymbol{\psi}^{a_{\ell k}}_{\ell,i}(\theta)}{\sum_{\theta'\in\Theta}\prod_{\ell\in\mathcal{N}_k}\boldsymbol{\psi}^{a_{\ell k}}_{\ell,i}(\theta')}, \quad k\in\mathcal{N}. \label{eq:combine}
    \end{align}
    \noindent The term in the denominator in~(\ref{eq:combine}) is used to normalize $\bmu_{k,i}(\theta)$ to a probability mass function. Once~(\ref{eq:combine}) is performed, the true state $\theta^{\star}$ can be estimated by agent $k$ at time $i$ using the maximum a-posteriori construction over either the private or public beliefs, for example:
    \begin{align}
        \widehat{\boldsymbol{\theta}}_{k,i} \triangleq \arg\max_{\theta\in\Theta}\bpsi_{k,i}(\theta).
        \label{eq:truestate_est0}
    \end{align}
    It was shown in~\cite{bordignon2020adaptive, shumovskaia2022explainability} that this social learning scheme has a powerful performance guarantee. Specifically, the probability of error goes to zero for both the private and public beliefs as the step-size $\delta$ approaches zero, namely,
    \begin{align}
        &\lim_{\delta \rightarrow 0}\lim_{i\rightarrow\infty} \mathbb P (\arg\max_{\theta\in\Theta}\boldsymbol{\mu}_{k,i}(\theta) \neq \theta^\star) = 0, \; \forall k \in \mathcal N
    \end{align}
    and
    \begin{align}
        \lim_{\delta \rightarrow 0}\lim_{i\rightarrow\infty} \mathbb P (\arg\max_{\theta\in\Theta}\boldsymbol{\psi}_{k,i}(\theta) \neq \theta^\star) = 0, \; \forall k \in \mathcal N.
        \label{eq:theta_learning}
    \end{align}
    Thus, the agents' confidence on hypothesis $\theta^\star$ being the true hypothesis converges to one. In other words, the agents are able to learn the truth eventually.
    
    Following~\cite{shumovskaia2022explainability}, we introduce two matrices $\blambda_i$ and $\bL_i$ in order to represent the recursions (\ref{eq:adapt_adaptive})--(\ref{eq:combine}) in a more compact matrix form as follows:
    \begin{align}
        \blambda_i=(1-\delta)A_\star^\bT\blambda_{i-1}+\delta\bL_i.
        \label{eq:recursion_adaptive}
    \end{align}
    The matrices are of size $|\mathcal N| \times (|\Theta|-1)$, and their entries are log-belief and log-likelihood ratios and given by:
    \begin{align}
        &[\boldsymbol{\Lambda}_{i}]_{k,j} \triangleq\log\frac{\boldsymbol{\psi}_{k,i}(\theta_0)}{\boldsymbol{\psi}_{k,i}(\theta_j)}
        \label{eq:lambda}\\
        &[\boldsymbol{\mathcal{L}}_{i}]_{k,j}\triangleq\log\frac{L_k(\boldsymbol{\bzeta}_{k,i}\vert\theta_0)}{L_k(\boldsymbol{\bzeta}_{k,i}\vert\theta_j)}
        \label{eq:loglikelihood}
    \end{align}
    where the reference state $\theta_0\in\Theta$ can be chosen at will by the designer. 
    
    The matrices $\bL_i$ are i.i.d. over time due to the i.i.d. assumption on the observations. Again, following similar previous approaches~\cite{bordignon2020adaptive, shumovskaia2022explainability}, we introduce the following condition on the higher-order moments of $\bL_i$.
    \begin{Asm}[\bf{Positive-definite covariance matrix}]
        \label{asm:loglikelihoods_positivedefinite}
        The covariance matrix $\R_{\bL}$ is uniformly positive-definite for all $i \geq 0$, i.e., there exists $\tau > 0$ such that:
        \begin{align}
            \R_{\bL} \triangleq \bE \left(\bL_i - \bE \bL_i \right)\left(\bL_i - \bE\bL_i\right)^\bT \geq \tau I.
        \end{align}
        \qedsymb
    \end{Asm}
    
    Iterating~(\ref{eq:recursion_adaptive}), we find that
    \begin{align}
        \blambda_i = \left(1-\delta\right)^i \left(A_\star^i\right)^\bT\blambda_0 + \delta \sum_{t=0}^{i-1} \left(1-\delta\right)^{t}(A_\star^{t})^\bT\bL_{i-t}.
        \label{eq:recursion_extended}
    \end{align}
    For large $i$, it can be shown that $\blambda_i$ converges in distribution to a limit value given by~\cite[Lemma 1]{shumovskaia2022explainability}:
    \begin{align}
        \blambda_i \xrightarrow{d} \blambda \triangleq \delta \sum_{t=0}^\infty \left(1-\delta\right)^{t}(A_\star^{t})^\bT\bL_t.
        \label{eq:blambda_lim_}
    \end{align}
    For further analysis, we introduce the following useful result for log-beliefs ratios. The standard laws of large numbers cannot be applied directly to the expression for $\blambda_i$ in~(\ref{eq:recursion_extended}) since there are dependencies between the random variables.
    \begin{Lem}[\bf{Law of large numbers for log-belief ratios}]
        \label{lemma:mean}
        After a sufficient number of iterations $i$ (i.e., for $i > M \gg 1$), the average of the log-belief matrix converges in probability as follows:
        \begin{align}
            \frac 1M \sum_{j=i-M}^{i-1} \blambda_j \xrightarrow{M\rightarrow\infty}  \bE \blambda.
        \end{align}
    \end{Lem}
    \begin{prf}
        See Appendix \ref{apx:mean}.\\
    \end{prf}

\section{Inverse Learning From Public Beliefs
}\label{sec:algorithm}
    In this section, we introduce an algorithm for learning the graph combination matrix $A_{\star}$ from observations of the public beliefs, as well as for assessing the level of informativeness of the various agents. 
    \subsection{Problem Statement}
        The data available from the social network might be limited for various reasons, including privacy.
        Therefore, in this work, we assume that we can only observe the evolution of the public beliefs over time:
        \begin{align}
            \left\{ \bpsi_{k,i}(\theta)\right\}_{i\gg 1},\;\;\forall k \in \mathcal N.
            \label{eq:sequence}
        \end{align}
        This assumption is motivated by the fact that agents \textit{share} their intermediate beliefs computed by~({\ref{eq:adapt_adaptive}}) during  the collaborative process, in contrast to the private beliefs in~(\ref{eq:combine}).
        Here, by $i\gg 1$ we underline that we are observing $\blambda_i$ after a sufficient amount of iterations, i.e., after $\bm{\Lambda}_i$ has reached its steady-state distribution~(\ref{eq:blambda_lim_}).
        
        A good illustration for this setting is the social network of Twitter users. 
        From each post (or tweet) that a user posts to their followers, we can extract an intermediate belief $\bpsi_{k,i}$ based on sentiment analysis, a.k.a., opinion mining. 
        After that, each user $k$ reads the posts of its followers and constructs the private belief $\bmu_{k,i}$ according to~(\ref{eq:combine}). 
        
        Next, we will show that by observing public beliefs, we can recover many useful network properties such as (i) the combination matrix, which determines the confidence levels that agents have about each other, (ii) the KL-divergences $D_{\textrm{KL}}\left(L_k(\theta^\star)||L_k(\theta)\right)$, which assess the capacity of each agent to distinguish the true hypothesis from the other possibilities, (iii) the pairwise  influences of agents on each other, and the (iv) global influencers across the network. 
        We allow both the network and the true state $\theta^\star$ to drift over time, and the algorithm will be able to track these changes too.
    
    \subsection{Algorithm Development}
        The previous work on learning the combination matrix $A_{\star}$ in~{\cite{shumovskaia2022explainability}}  assumes that the expected log-likelihood matrix $\bar\cL \triangleq \bE \mathcal \bL_i$ is known. It can be verified that 
        \begin{align}
            &\;[\bar \cL]_{k,j} \triangleq [\bE \bL_i]_{k,j} \nonumber\\
            =&\; D_{\textrm{KL}} (L_k(\theta^\star)||L_k(\theta_j)) - D_{\textrm{KL}} (L_k(\theta^\star)||L_k(\theta_0)).
            \label{eq:barL}
        \end{align}
        Using this knowledge along with~(\ref{eq:recursion_adaptive}), the following objective function was then minimized to learn $A_\star$:
        \begin{align}\label{eq:cost_function_original} 
            &Q(A; \blambda_i, \blambda_{i-1}) 
            \triangleq \frac{1}{2} \| \boldsymbol{\Lambda}_i -(1-\delta)A^{\mathsf{T}}\boldsymbol{\Lambda}_{i-1} - \delta \bar \cL \|_{\textrm F}^2
        \end{align}
        in terms of the squared Frobenius norm. In this work, we do not assume that $\bar \cL$ is known beforehand, and will instead estimate $\bar \cL$ at each iteration $i$ by using
        \begin{align}
            \widehat \bL_{i-1}(A) = \frac{1}{\delta M} \sum_{j=i-M}^{i-1} \left(\blambda_j - (1-\delta)A^\bT \blambda_{j-1}\right),
            \label{eq:loglikelihood_estimator}
        \end{align}
        where $A$ will be the estimate that is available for $A_{\star}$ at that point in time. This step allows us to keep any information about the privately received data $\bzeta_{k,i}$ hidden from the algorithm, which makes potential applications more feasible. Accordingly, the cost function is replaced by:
        \begin{align}
            &\;\widehat Q(A; \blambda_i, \blambda_{i-1}, \widehat \bL_{i-1}) \nonumber\\
            &\;= \frac{1}{2} \| \boldsymbol{\Lambda}_i -(1-\delta)A^{\mathsf{T}}\boldsymbol{\Lambda}_{i-1} - \delta \widehat \bL_{i-1} \|_{\textrm F}^2\nonumber\\
            &\;=\frac{1}{2} \Bigg\| \blambda_i - \frac 1M \sum_{j=i-M}^{i-1}\blambda_{j} - (1-\delta) A^\bT \nonumber\\
            &\;\;\;\;\;\times \left(\blambda_{i-1} - \frac 1M \sum_{j=i-M}^{i-1}\blambda_{j-1}\right)\Bigg\|_{\textrm F}^2 \nonumber\\
            &\;= \frac 12  \|\bdelta_i - (1-\delta)A^\bT \bdelta_{i-1}\|_{\textrm F}^2
            \label{eq:cost_function}
        \end{align}
        where we introduced
        \begin{align}
            \bdelta_i \triangleq \blambda_i - \frac 1M \sum_{j=i-M}^{i-1} \blambda_j
            \label{eq:bdelta}
        \end{align}
        The corresponding risk function is then given by:
        \begin{align}
            J_i(A) \triangleq \bE \widehat Q(A; \blambda_i, \blambda_{i-1}, \widehat \bL_{i-1})
            \label{eq:risk}
        \end{align}
        We will show in Lemma~\ref{lemma:risk} that the unique minimizer of this risk function gets closer to the true combination matrix as $M$ grows.
        
        Now, assume that we observe a sequence of public beliefs~(\ref{eq:sequence}), therefore a sequence $\left\{ \blambda_i\right\}_{i\gg 1}$. We define the objective function as a sum over $N$ observed time indices $i\gg 1$:
        \begin{align}
        \label{eq:opt}
            &\min_{A} J(A) \triangleq \frac 1{N-M}\sum_{i} J_i(A)
        \end{align}
        To solve~(\ref{eq:opt}), we apply stochastic gradient descent (SGD) with constant step-size $\mu > 0$. At each iteration $i$, the estimate $\bA_i$ for the combination matrix is updated via:
        \begin{align}
            \bA_i 
            =&\; \bA_{i-1}
            + \mu(1-\delta)\bdelta_{i-1} \nonumber\\
            &\times\left(\blambda_i^\bT - (1-\delta) \blambda_{i-1}^\bT \bA_{i-1} - \delta \widehat \bL_{i-1}^\bT\right)
            \label{eq:descent}
        \end{align}
        We sample the observations in the direct order $i, i+1, i+2\dots$. This online nature of the algorithm, as well as the use of a constant step-size instead of a vanishing step-size, will allow the algorithm to track changes in ${A_\star}^{\textrm{\footnotemark}}$. \footnotetext{In our algorithm, we assume deterministic $A_\star$. However, recursion~(\ref{eq:descent}) allows adapting to changes in $A_\star$.}We list the procedure in Algorithm~\ref{alg}.
        
        \begin{algorithm}
            \KwData{
                At each time $i$:
                \begin{align*}
                    \left\{ \bpsi_{k,i}(\theta)\right\}_{k\in \mathcal N}, \delta
                \end{align*}
            }
            \KwResult{Estimated combination matrix $\bA_N$, \\
            $\;\;\;\;\;\;\;\;\;\;\;\;$estimated expected log-likelihood ratios $\widehat{\bL}_N$.\\
            }
            initialize $\bA_0$, $\widehat \bL_0$\\
            \Repeat{sufficient convergence}{
                Compute matrices $\blambda_i$:\\
                \For{$k\in\mathcal N$, $j=1,\dots,|\Theta|$}{
                    \begin{flalign*}
                        &[\boldsymbol{\Lambda}_{i}]_{k,j} = \log\frac{\boldsymbol{\psi}_{k,i}(\theta_0)}{\boldsymbol{\psi}_{k,i}(\theta_j)}&&
                    \end{flalign*}
                }
                \noindent Combination matrix update:
                \begin{flalign*}
                    &\boldsymbol{A}_i =  \boldsymbol{A}_{i-1} + \mu(1-\delta)\left(\blambda_{i-1} - \frac 1M \sum_{j=i-M}^{i-1} \blambda_{j-1}\right) \nonumber&&\\
                    &\textrm{ }\textrm{ }\textrm{ }\textrm{ }\textrm{ }\textrm{ }\times\left(\blambda_i^{\mathsf{T}} -  (1-\delta)\blambda_{i-1}^{\mathsf{T}}\boldsymbol{A}_{i-1} - \delta \widehat \bL_{i-1}^\bT \right).&&
                \end{flalign*}
                \noindent Log-likelihoods matrix update:
                \begin{flalign*}
                    &\widehat \bL_i = \frac 1{\delta M} \sum_{j=i-M+1}^{i}\left(\blambda_j - (1-\delta)\bA_{i}^\bT \blambda_{j-1}\right)&&
                \end{flalign*}
                \begin{flalign*}
                    &i = i+1&&
                \end{flalign*}
            }
            \caption{Graph social learning (GSL)}
            \label{alg}
        \end{algorithm}

\section{Global Influence Identification}
\label{sec:influence_identification}
    In this section, we establish a strong connection between the probability of error for truth learning and the \textit{network divergence}. The network divergence is defined in terms of the Perron eigenvector of $A_\star$, and the KL-divergences between the likelihoods:
    \begin{align}
        K(\theta^\star,\theta) \triangleq \sum_{k\in\mathcal N} u_k D_{\textrm{KL}} (L_k(\theta^\star)||L_k(\theta)) > 0.
        \label{eq:global_divergence}
    \end{align}
    
    \noindent Note that this quantity is a function of the hypothesis $\theta$. As remarked before~(\ref{eq:truestate_est0}), the true state estimator can be deduced from the public beliefs.
    The probability of error is defined as the probability of selecting a wrong hypothesis $\theta\neq\theta^\star$:
    \begin{align}
        p_{k,i} \triangleq \mathbb P \left( \arg\max_{\theta \in \Theta}  \bpsi_{k,i}(\theta) \neq \theta^\star \right). 
        \label{eq:prob_error}
    \end{align}
    Note that if $\theta^\star$ does not maximize a public belief, then there exists at least one $\theta\neq\theta^\star$ such that
    \begin{align}
        \log \frac{\bpsi_{k,i}(\theta^\star)}{\bpsi_{k,i}(\theta)} \leq 0.
        \label{eq:logratio}
    \end{align}
    Thus, we can equally define the probability of error $p_{k,i}$ as:
    \begin{align}
         p_{k,i} = \mathbb P \left( \exists \theta \neq \theta^\star \colon \log \frac{\bpsi_{k,i}(\theta^\star)}{\bpsi_{k,i}(\theta)} \leq 0 \right).
         \label{eq:perror}
    \end{align}
    For this section alone, we will additionally assume that the observations $\{\bzeta_{k,i}\}$ are independent over space (and not only over time). Now, we know from~\cite[Theorem 3]{bordignon2020adaptive} that each random variable (for any $k$) of the form~(\ref{eq:logratio}) can be approximated by a Gaussian random variable in the steady state with the following moments:
    \begin{align}
        \log \frac{\bpsi_{k,i}(\theta^\star)}{\bpsi_{k,i}(\theta)} \approx \mathcal G \Big(K\left(\theta^\star, \theta\right) + O\left(\delta\right), \delta C + O\left(\delta^2\right)\Big)
        \label{eq:gv}
    \end{align}
    for some finite and constant covariance matrix, $C$. Thus, the probability of error~(\ref{eq:perror}) becomes the probability of the Gaussian random variable~(\ref{eq:gv}) assuming negative values for at least one $\theta\in\Theta$.
    This Gaussian random variable concentrates around its mean (i.e., the network divergence in (\ref{eq:global_divergence})), which is positive. The larger network divergence is, the smaller the probability of error for each individual agent will be.
    
    If we examine expression (\ref{eq:global_divergence}) for the network divergence, we observe that each individual agent $k$ contributes with a term of the form
    \begin{align}
        K_k(\theta^\star,\theta) = u_k D_{\textrm{KL}} (L_k(\theta^\star)||L_k(\theta))
        \label{eq:network_divergence_k}
    \end{align}
    which is scaled by the Perron entry $u_k$. This entry reflects the centrality of agent $k$,  and serves as a measure of how well it is connected to other nodes in the network. 
    The larger the value of (\ref{eq:network_divergence_k}) is, the stronger the contribution of this agent will be towards moving the network away from an erroneous decision. We therefore say that (\ref{eq:network_divergence_k}) helps convey the amount of information that agent $k$ has about $\theta$ disagreeing with $\theta^{\star}$. Since (\ref{eq:network_divergence_k}) is a function of $\theta$, we can define the level of informativeness of agent $k$ to the learning process by considering the aggregate of its contributions for all $\theta$, which we denote by 
    \begin{align}
        I_k \triangleq \sum_{\theta\in\Theta} K_k(\theta^\star,\theta) = u_k \sum_{\theta\in\Theta} D_{\textrm{KL}} (L_k(\theta^\star)||L_k(\theta)).
        \label{eq:influence}
    \end{align}
   This quantity serves as a measure of \textit{influence}, since agents with large $I_k$ contribute the most to learning the truth by the network.

    In what follows, we describe how to estimate the quantities $I_k$ by the learning algorithm. First, to obtain the Perron eigenvector for $\bA_i$, we need to normalize any of its eigenvectors corresponding to the eigenvalue at 1. Subsequently, we identify $\widehat{\boldsymbol\theta}_{k,i}$ that maximizes $\bpsi_{k,i}(\theta)$.
    From~(\ref{eq:theta_learning}) we know that $\widehat{\boldsymbol{\theta}}_{k,i}$ tends to $\theta^\star$ almost surely as $i\rightarrow\infty$ and $\delta \rightarrow 0$. After a sufficient number of iterations $i_N$ of the Algorithm~\ref{alg}, we let $j'$ denote the index within the hypothesis set $\Theta$ that corresponds to:
    \begin{align}
        \widehat {\boldsymbol\theta}_{j'} =     \arg\max_{\theta\in\Theta} \bpsi_{k,i_N}(\theta).
        \label{eq:psi_estimate_}
    \end{align}
    Returning to~(\ref{eq:barL}), we can then approximate the KL-divergences by
    \begin{align}
        \label{eq:dkl1}
        &D_{\textrm{KL}}\left(L_k(\theta^\star)||L_k(\theta_0)\right) \approx -[\widehat \bL_{i_N}]_{k,j'},\\
        \label{eq:dkl2}
        &D_{\textrm{KL}}\left(L_k(\theta^\star)||L_k(\theta_j)\right) \approx [\widehat \bL_{i_N}]_{k,j} + [\widehat \bL_{i_N}]_{k,j'},
    \end{align}
    where $\widehat \bL_{i_N}$ is an estimate for $\bar\cL$.
    
    To conclude, the sequence of public beliefs contains rich information about the network. Using the GSL algorithm, we are not only able to identify the graph topology, but can also find answers to the explainability question: which agents were the most responsible (or the main drivers) for the overall network learning process?
        
\section{Theoretical Results}\label{sec:theory}
    First, we establish some useful properties of the risk function~(\ref{eq:risk}).
    \begin{Lem}[\bf{Risk function properties}]
        \label{lemma:risk}
        After sufficient number of iterations $i$, the risk function $J_i(A)$ is strongly convex and has Lipschitz gradient with constants $\nu_i$ and $\kappa_i$ given by:
        \begin{align}
            \nu_i &\;\triangleq (1-\delta)^2\lambda_{\min}\left(\bE \bdelta_{i-1}\bdelta_{i-1}^\bT \right) \nonumber\\
            &\; \geq \tau\delta^2(1-\delta)^2 + O\left(1/\sqrt M\right)
            \label{eq:nu}\\
            \kappa_i &\;\triangleq (1-\delta)^2\lambda_{\max}\left(\bE \bdelta_{i-1}\bdelta_{i-1}^\bT \right)\nonumber\\
            &\; \geq \tau\delta^2(1-\delta)^2 + O\left(1/\sqrt M\right)
            \label{eq:kappa}
        \end{align}
        where $\lambda_{\min}(\cdot)$ and $\lambda_{\max}(\cdot)$ are the minimum and maximum eigenvalues. In other words, it holds that:
        \begin{align}
            \nu_i I \leq \nabla_A^2 J_i(A) \leq \kappa_i I.
        \end{align}
        Moreover, the difference between the true combination matrix $A_\star$ and the unique minimizer $A_{\min, i}$ of $J_i(A)$ is on the order of:
        \begin{align}
            \left\|A_{\min, i} - A_\star \right\|_{\rm F}^2 \leq O\left( {1}/ \delta^2 {M^2}\right)
            \label{eq:A_diff}
        \end{align}
    \end{Lem}
    \begin{prf}
        See Appendix \ref{apx:risk}.\\
    \end{prf}
    \noindent Result~(\ref{eq:A_diff}) states that the gap between $A_\star$ and the unique minimizer of $J_i(A)$ is negligible as $M \gg 1/\delta $ grows. 
    
    To investigate the steady-state performance of recursion~(\ref{eq:descent}), we adopt the following independence assumption, which is typical in the study of adaptive systems~\cite{sayed2011adaptive, Sayed_2014, sayed2013diffusion}.
    \begin{Asm}[\bf{Separation principle}]
        \label{asm:independence}
        We denote the estimation error by $\widetilde\bA_i \triangleq A_\star - \bA_i $, and assume the step-size $\mu$ is small enough to allow $\|\widetilde\bA_i\|_{\textrm{\textup{F}}}^2$ to attain a steady-state distribution. The separation principle states that the error $\widetilde\bA_i$ is independent of the observations $\blambda_i,\dots,\blambda_{i-M}$, conditioned on the history of previous observations.
        \qedsymb
    \end{Asm}
    \noindent The following theorem shows that in steady-state, the mean-squared error is $O(\mu) + O(1/\delta^3 M^2)$ in expectation.
    \begin{Thm}[\bf{Steady-state performance}]
        The mean-square deviation (MSD) converges exponentially fast with asymptotic convergence rate:
        \begin{align}
            \alpha \triangleq&\; 1 - \mu (2 \nu + O(\delta^3)) + O(\mu^2)
            \label{eq:conv_rate}
        \end{align}
        where
        \begin{align}
            \nu &\;\triangleq (1-\delta)^2\lambda_{\min}\left(\lim_{i\rightarrow\infty}\bE \bdelta_{i-1}\bdelta_{i-1}^\bT \right)
        \end{align}
        In the limit, the MSD satisfies:
        \begin{align}
            \limsup_{i\rightarrow\infty} \bE \|\widetilde{\bA}_{i}\|_{\textrm{\textup{F}}}^2 \leq 
            O(\mu) + O(1/\delta^3 M^2)
            \label{eq:msd_lim}
        \end{align}
        \label{thm:conv}
    \end{Thm}
    \begin{prf}
        See Appendix \ref{apx:conv}.\\
    \end{prf}
    \noindent Naturally, the convergence rate for learning the combination matrix is dependent on the strong convexity constant, $\nu$. Usually, the limiting MSD expression is of the form  $O\left(\mu\right)$, so that we can reduce the deviation by using arbitrary small step-size $\mu$. In our case, we also have the additional term  $O\left( 1 /\delta^3 M^2\right)$, which is due to the difference~(\ref{eq:A_diff}) between the unique minimizer and the true combination matrix. We can control the number of samples $M$. Note that by selecting $M=O(1/\sqrt \mu)$, the MSD in~(\ref{eq:msd_lim}) becomes $O(\mu)$.
    
    Finally, we study how well the algorithm approximates $\bar \cL$.
    \begin{Thm}[\bf{Steady-state log-likelihood learning}]
        The MSD converges exponentially fast with
        \begin{align}
            &\;\limsup_{i\rightarrow\infty} \bE \|\widehat \bL_{i} - \bar \cL\|_{\textrm{\textup{F}}}^2 \nonumber\\
            &\;\leq \frac 1M \textup{Tr} \left(\R_{\bL}\right) + O(\mu/\delta^2) + O\left( 1 / \delta^5 M^2 \right)
        \end{align}
        where
        \begin{align}
            \R_{\bL} = \bE \left(\bL_i - \bar \cL\right)\left(\bL_i - \bar \cL\right)^\bT
        \end{align}
        is independent of $i$ due to i.i.d. observations.
        \label{thm:conv2}
    \end{Thm}
    \begin{prf}
        See Appendix \ref{apx:conv2}.\\
    \end{prf}
    \noindent Since we simultaneously learn the combination matrix $A_\star$ and $\bar\cL$, the limiting MSD for $\bar \cL$ has similar expression to Theorem~\ref{thm:conv}. We would like to note that usually, the hyper parameter $\delta$ is a network property we don not have a direct influence on. Thus, the final MSD is controlled by the step-size $\mu$ and it decreases as we use higher number of samples $M$. 
    
\section{Computer Simulations}\label{sec:experiments}
    In this section, we illustrate how well the proposed algorithm is able to identify the true combination matrix $A_\star$ and the expected log-likelihood matrix $\bar\cL$. We also experiment with different $M$ in~(\ref{eq:loglikelihood_estimator}) to see how the convergence rate changes. Additionally, we will see how well the learned combination matrix and KL-divergences identify the influences~(\ref{eq:influence}).
    
\begin{figure}
        \begin{subfigure}[b]{0.24\textwidth}
            \includegraphics[width=\linewidth]{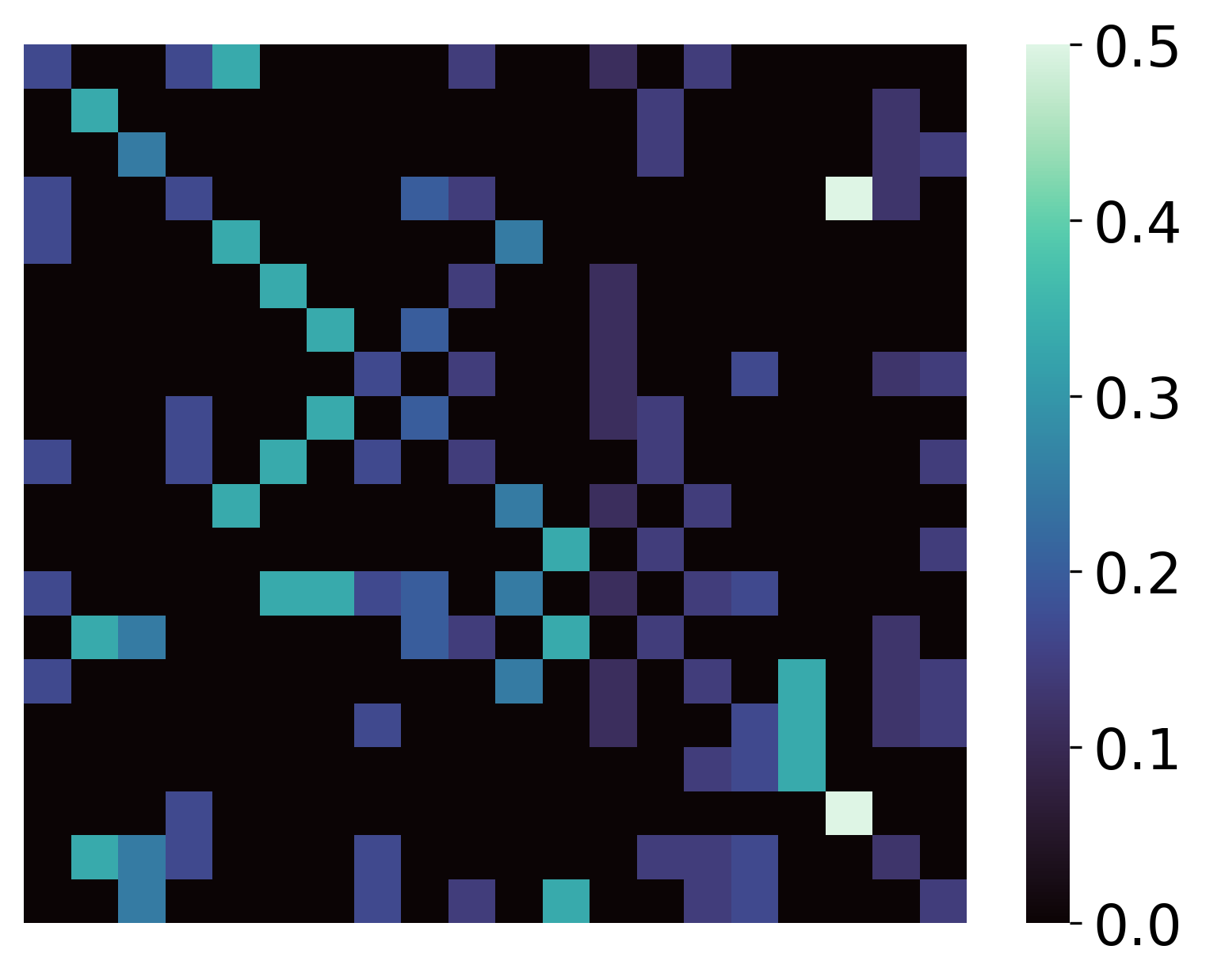}
            \caption{True combination matrix.} \label{fig:adj_true}
        \end{subfigure}\hspace*{\fill}
        \begin{subfigure}[b]{0.24\textwidth}
            \includegraphics[width=\linewidth]{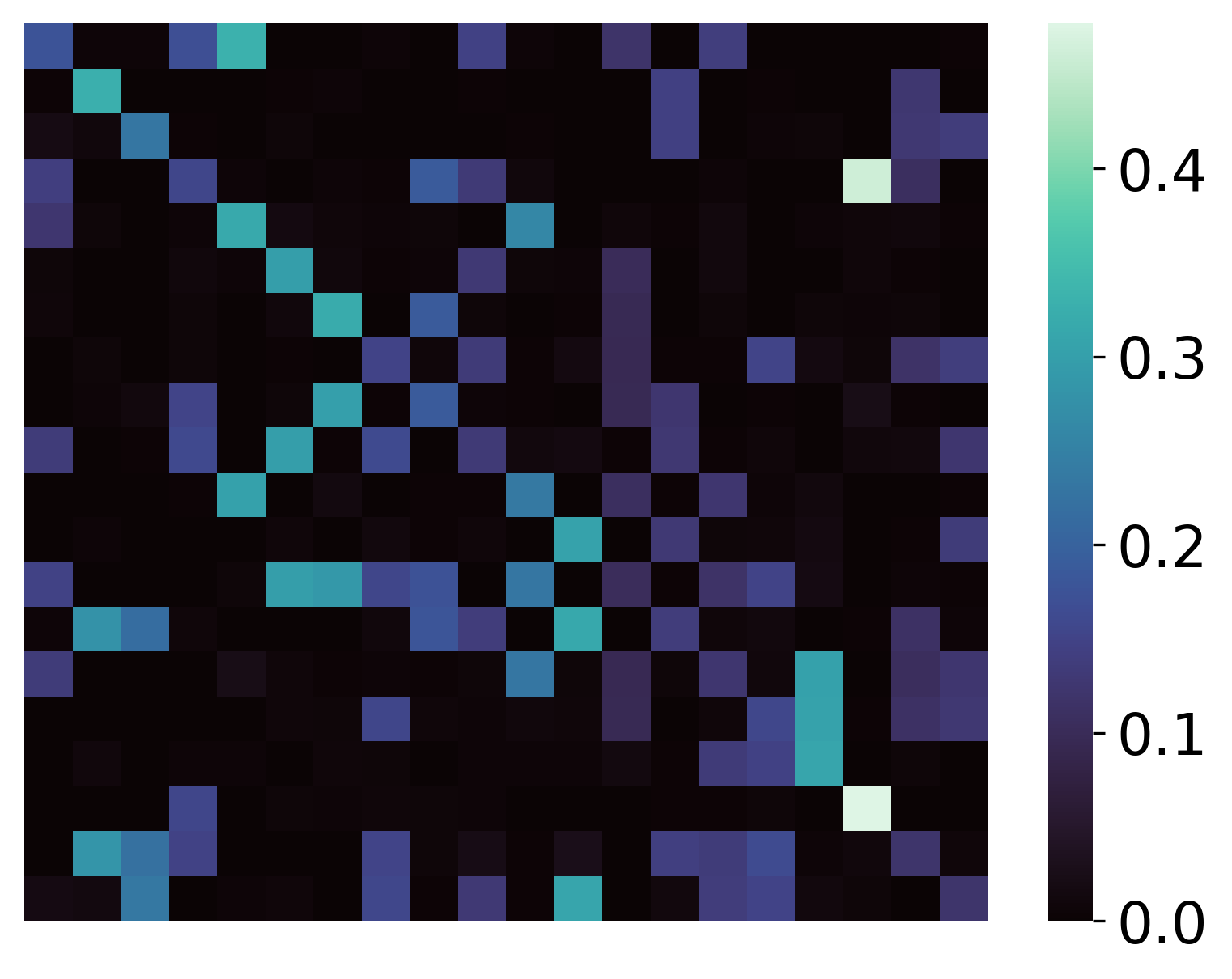}
            \caption{Learned combination matrix.} \label{fig:adj_learned}
        \end{subfigure}
        
    \caption{True combination matrix and the learned matrix using the GSL algorithm with $M = 50$.}
    \label{fig:adj}
    \end{figure}

    We generate a graph with $|\mathcal N| = 20$ agents according to the Erdos-Renyi model with an edge probability of $p=0.2$. We set the adaptation hyperparameter to $\delta = 0.05$. Then, we generate the combination weights (see Fig.~\ref{fig:adj_true}
    ) with uniform weights in the column, such that the resulting matrix is left-stochastic. We consider $|\Theta|=5$ states, where the likelihood models $L_k(\theta)$ for each agent $k\in\mathcal N$ are assumed to follow a  binomial distribution with randomly generated parameters (for details, see Appendix~{\ref{apx:generator}}). We generate likelihood models such that we observe only 3 agents with high informativeness. Later, we illustrate that the algorithm allows for identifying these agents.
    
\begin{figure}
        \centering
        \includegraphics[width=0.95\linewidth]{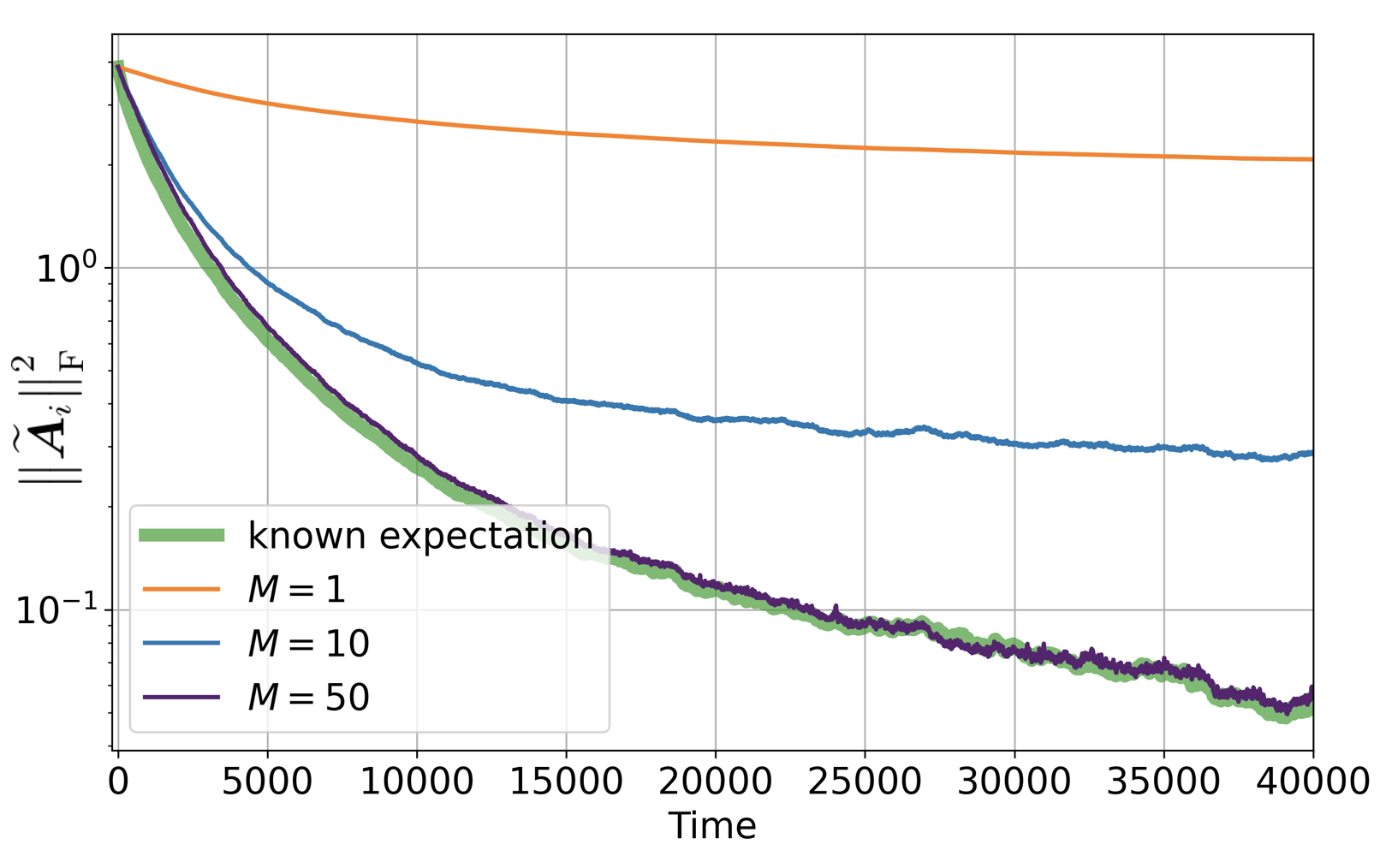}
        \caption{Algorithm performance when $\bar \cL$ is known and when it is estimated by~(\ref{eq:loglikelihood_estimator}) for different $M\in\{1,10,50\}$.}
        \label{fig:graph_error}
    \end{figure}

    First, we consider how well the combination matrix is learned for different $M \in \{1, 10, 50\}$. We additionally compare with~\cite{shumovskaia2022explainability}, where the expectation $\bar \cL$ was assumed to be known beforehand. For $M = 50$, we use $\mu=0.1$, for $M=10$, we use $\mu = 0.01$, and for $M=1$, we use $\mu=0.001$ for better convergence. In Fig.~\ref{fig:graph_error}, we plot the reconstruction error with respect to the iteration number:
    \begin{align}
        \|\widetilde{\bA}_{i}\|^2_{\textrm F} = \|\bA_i - A_\star\|_{\textrm F}^2
    \end{align}
    We notice that the higher $M$ improves the limiting MSD as reflected in Theorem~\ref{thm:conv}. 
    
\begin{figure}
        \centering
        \includegraphics[width=0.95\linewidth]{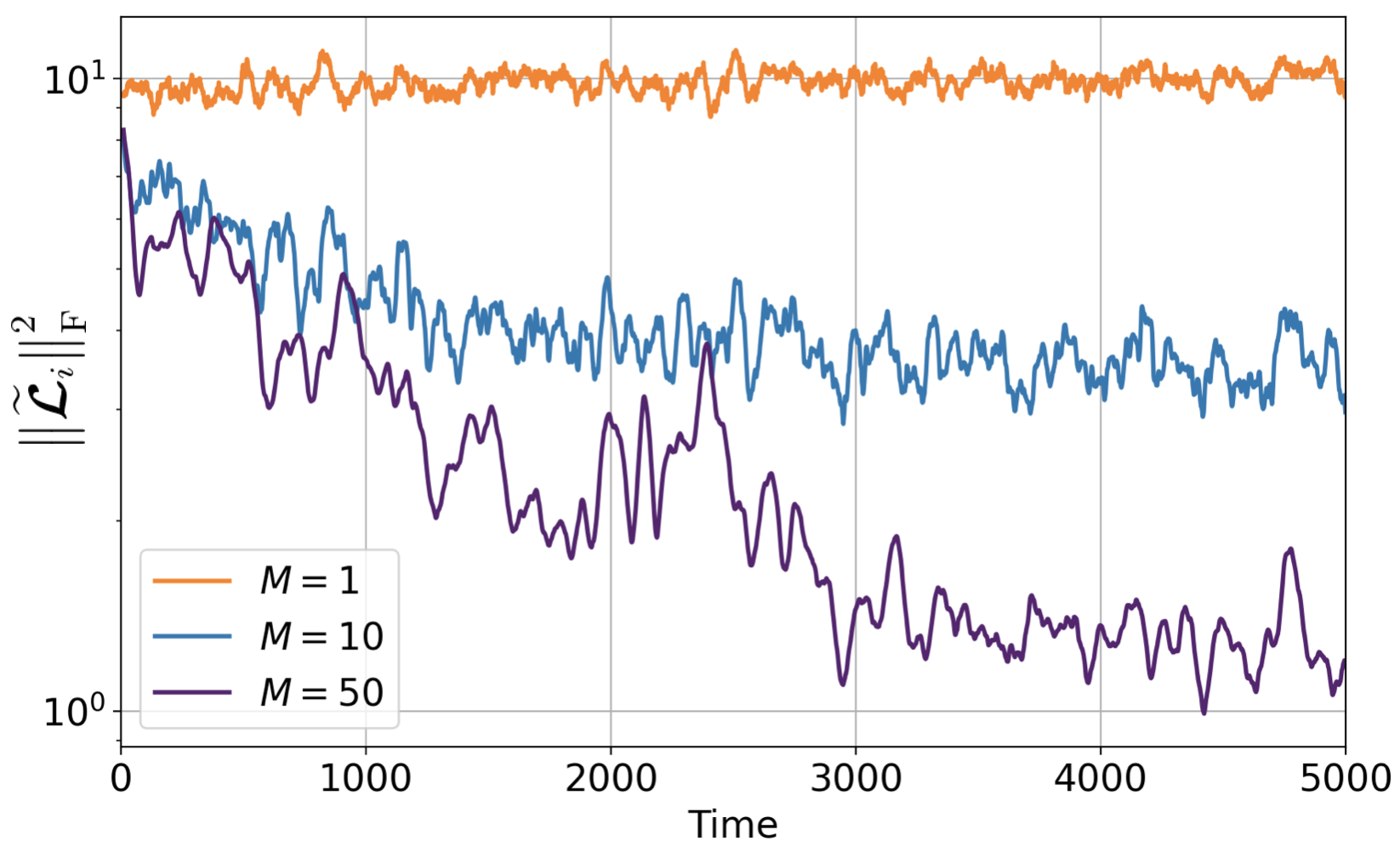}
        \caption{$\bar \cL$ reconstruction error estimated by~(\ref{eq:loglikelihood_estimator}) for different $M\in\{1,10,50\}$. To reduce the variance and therefore for better interpretability, we plot the rolling mean with window size equal to 50.}
        \label{fig:L_error}
    \end{figure}

    Next, we illustrate how well the learned $\widehat \bL_i$ approximates $\bar \cL$ for different $M \in \{1,10,50\}$. In Figure~\ref{fig:L_error}, we show how the reconstruction error evolves with $i$:
    \begin{align}
        \|\widetilde{\bL}_{i}\|_{\textrm F}^2 = \|\widehat \bL_i - \bar \cL\|_{\textrm F}^2.
    \end{align}
    We notice that with $M$ growing, we can approximate $\bar \cL$ more precisely, which aligns with the result of Theorem~\ref{thm:conv2}, and explains Figure~\ref{fig:graph_error}.
    
    We illustrate the recovered combination matrix in Fig.~\ref{fig:adj_learned}
    . 
    We use $M=50$ since it has better convergence. Comparing Fig.~\ref{fig:adj_true} and~\ref{fig:adj_learned}, we observe an almost perfect recovery of the true combination weights.  
    
\begin{figure}
        \centering
        \begin{subfigure}[b]{0.45\textwidth}
            \centering
            \subcaption{3 influential agents.}
            \label{fig:influences3}
            \includegraphics[width=\linewidth]{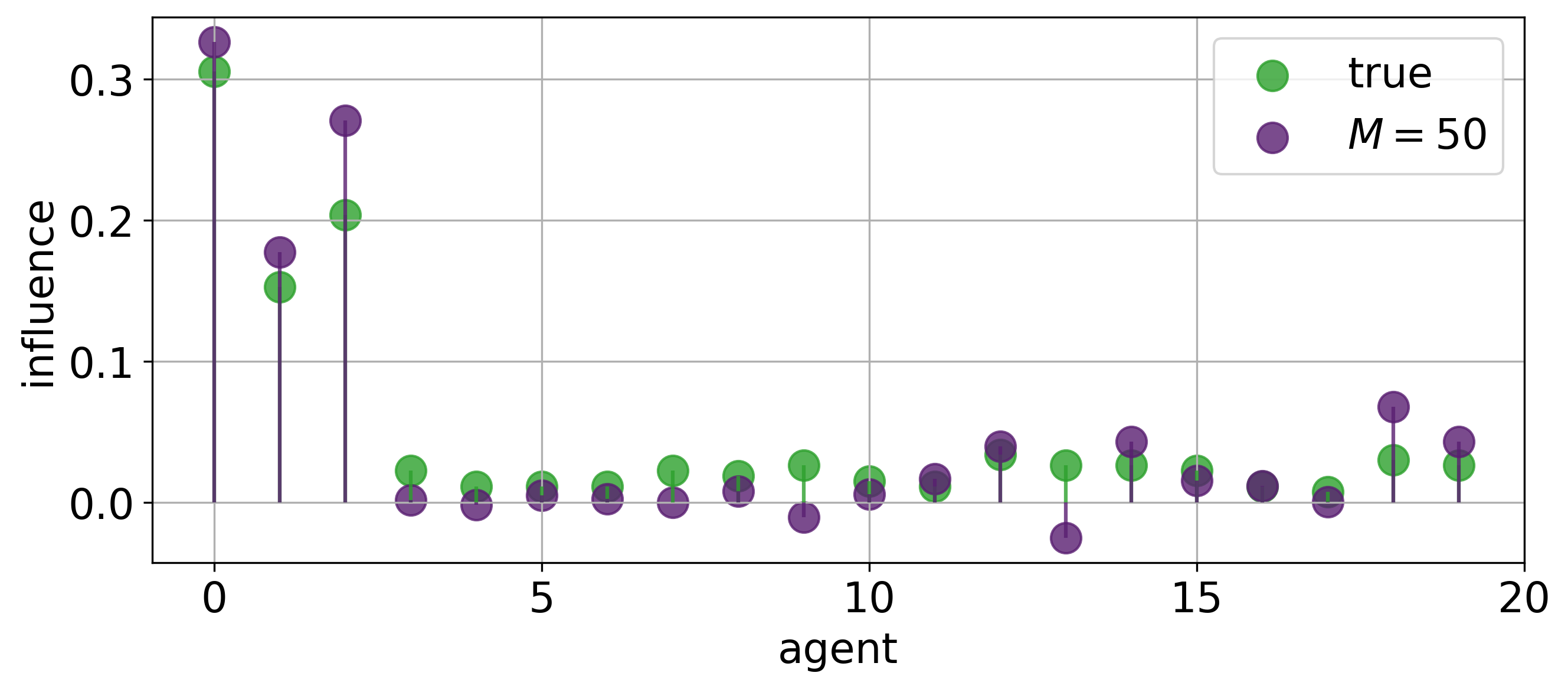}
        \end{subfigure}
        \vfill
        \begin{subfigure}[b]{0.45\textwidth}
            \centering 
            \subcaption{1 influential agent.}
            \label{fig:influences1}
            \includegraphics[width=\textwidth]{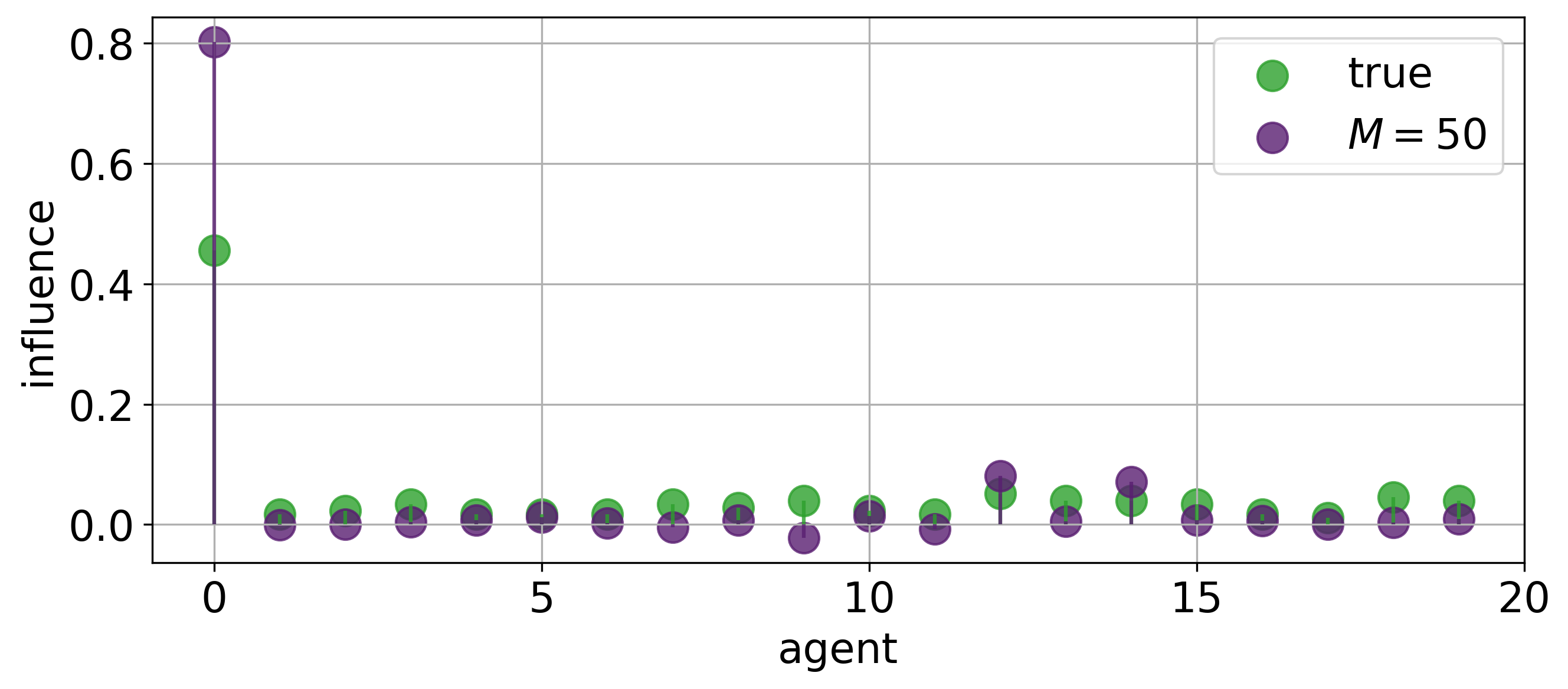}
        \end{subfigure}
        \caption{Agents' influences~(\ref{eq:influence}) based on the learned graph and KL-divergences.}
        \label{fig:influences}
    \end{figure}

    Figure~\ref{fig:influences} illustrates how well the learned KL-divergences and combination matrix can recover the global influences~(\ref{eq:influence}). For better interpretability, we normalize the values so that they add up to one. We see that for some agents, the algorithm does not perfectly recover these components, but yet allows us to identify that the first agents are driving the learning the most. This property allows us to search for agents that are the most contributing to learning the true state.
    
\begin{figure}
        \centering
        \includegraphics[width=0.95\linewidth]{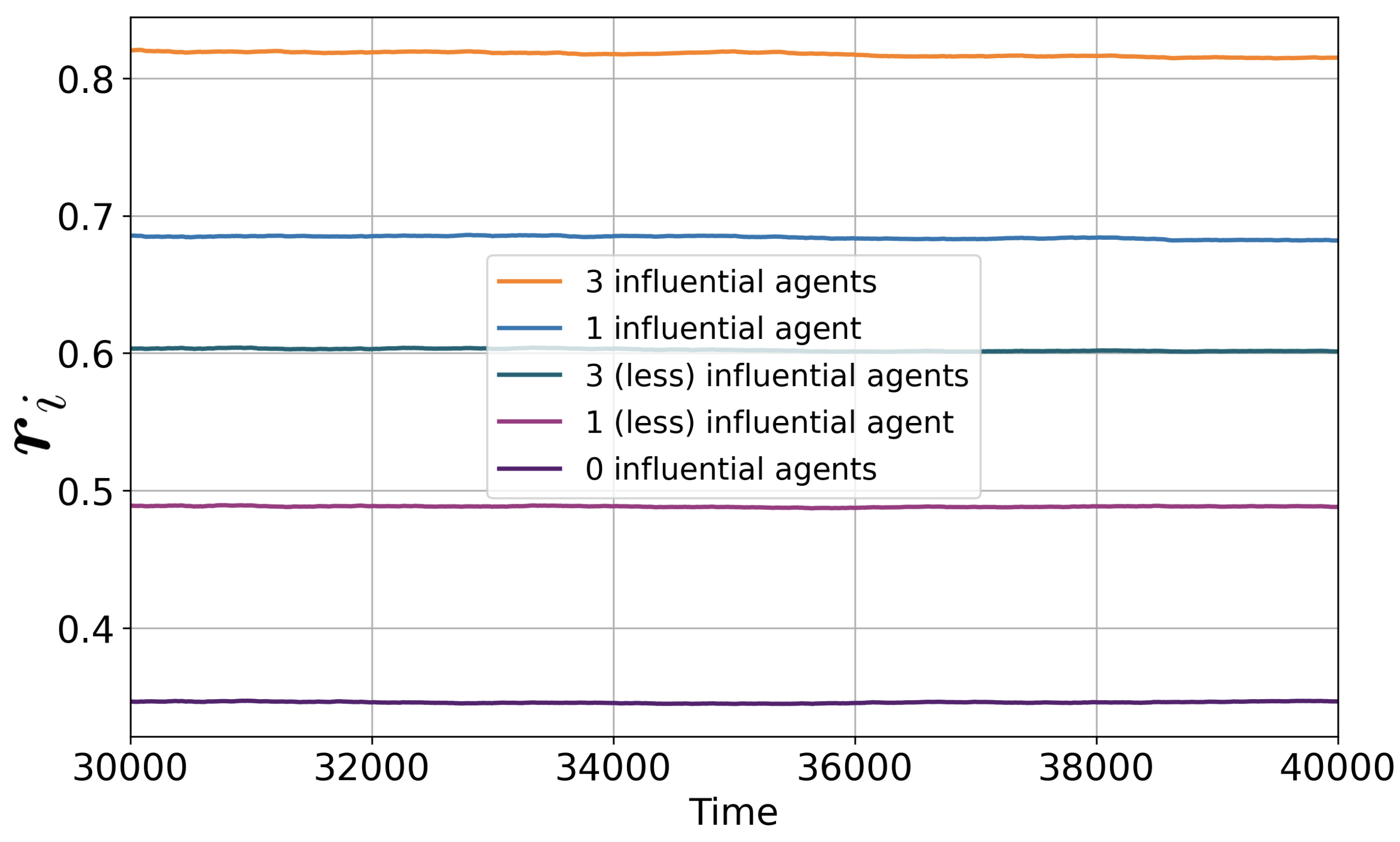}
        \caption{Rate of correctly classified truth for different models. Less influential agents denote agents with the same centrality as influential agents, but with smaller KL-divergence between states.}
        \label{fig:rate}
    \end{figure}

    In Fig.~\ref{fig:rate}, we illustrate the connection between the probability of error~(\ref{eq:prob_error}) and the presence of  agents with a significant contribution $I_k$~(\ref{eq:influence}) to the network divergence. We plot the rate of correctly classified states at each point $i$:
    \begin{align}
        \boldsymbol r_i = \frac 1i \sum_{t = 1}^i \mathbb I \{\widehat{\boldsymbol{\theta}}_i = \theta_\star \} 
    \end{align}
    We generate likelihood models according to Appendix~\ref{apx:generator}, where less (yet significantly) influential agents have smaller KL-divergence between likelihood models (lines 3 and 4 in the legend). Thus, we illustrate the value of the identified importance measure~(\ref{eq:influence}) and the role of ``informativeness" (or KL-divergences) of each agent: the presence of agents with high informativeness improves the quality of true state inference.

\begin{figure}
        \centering
        \begin{subfigure}[b]{0.45\textwidth}
            \centering
            \subcaption{Algorithm performance when the graph regularly changes its topology (every 1000 iterations). Every element of the adjacency matrix changes its state with probability $0.5\%$.}
            \label{fig:perturbe1}
            \includegraphics[width=\linewidth]{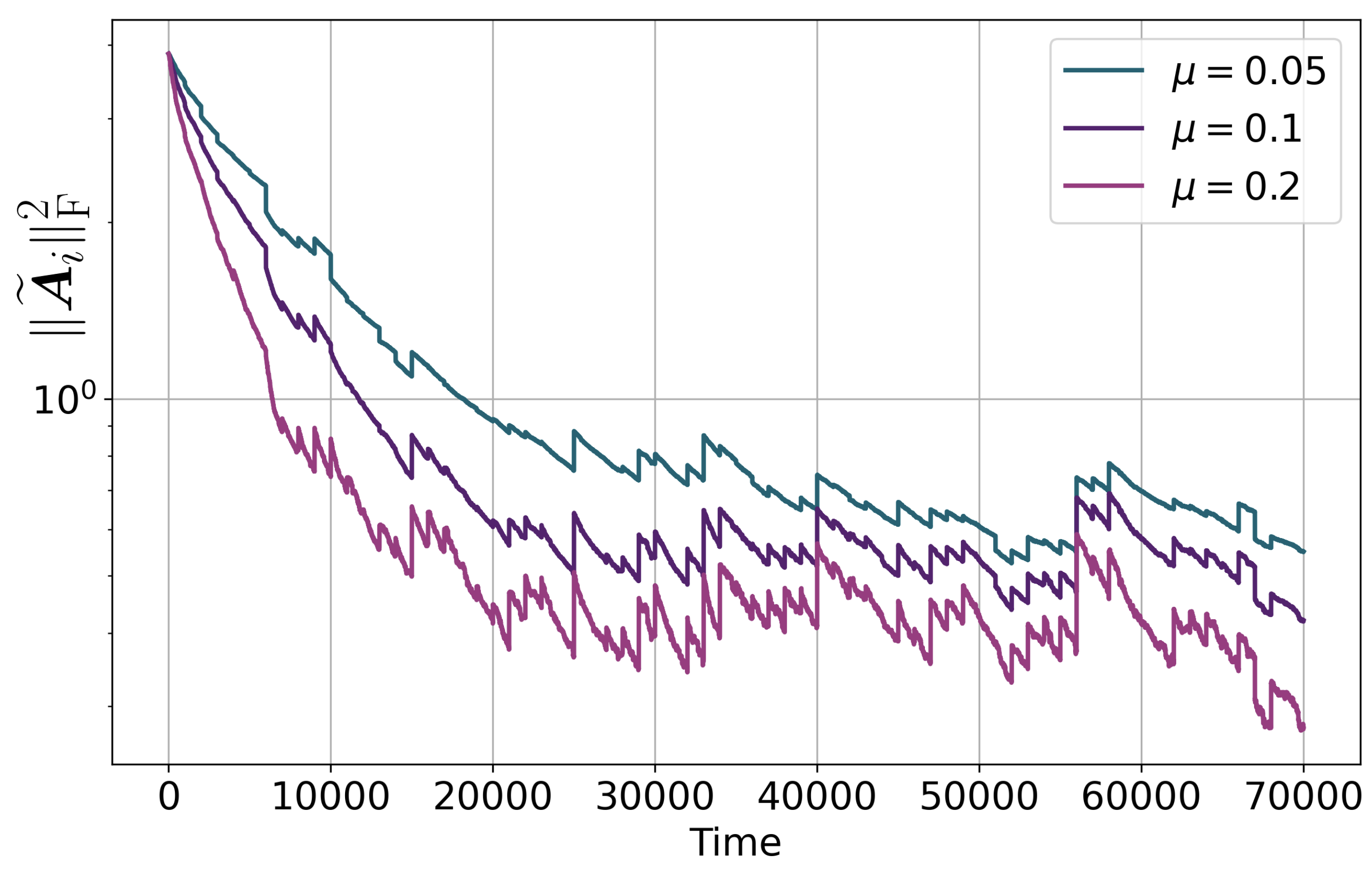}
        \end{subfigure}
        \vfill
        \begin{subfigure}[b]{0.45\textwidth}
            \centering 
            \subcaption{Algorithm performance when the true state $\theta^\star$ changes.}
            \label{fig:perturbe2}
            \includegraphics[width=\textwidth]{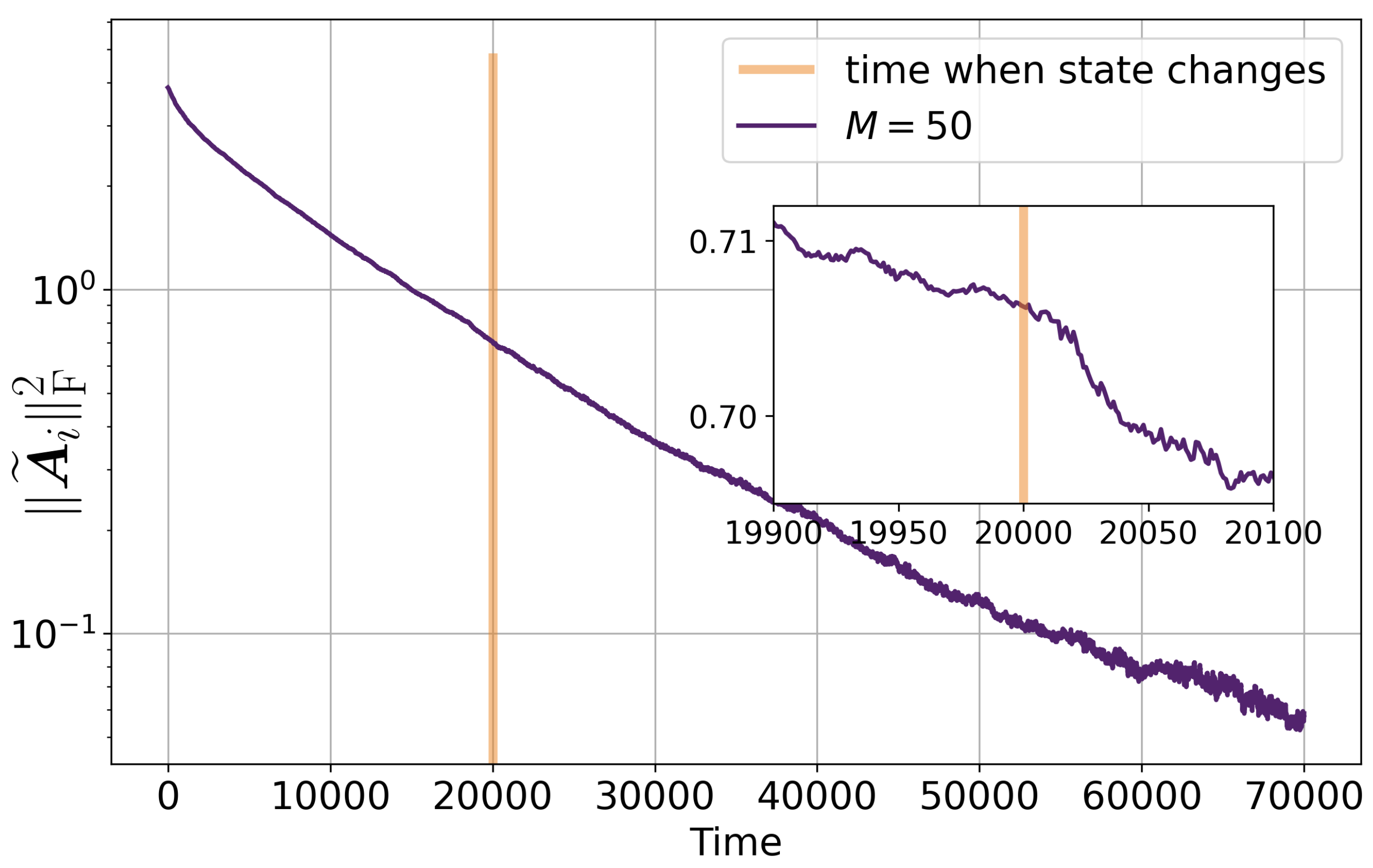}
        \end{subfigure}
        \caption{Algorithm performance under perturbations.}
        \label{fig:perturbe}
    \end{figure}

    Finally, we comment on the adaptation abilities of the proposed algorithm that results from its online nature~(\ref{eq:descent}). There is a trade-off between the speed of adaptation to changes in $A_\star$ or in $\theta^\star$ and the final MSD. Larger step-size $\mu$ leads to faster reaction by the algorithm to changes by the combination matrix, but enlarges the steady-state MSD. 
    However, an important point to note here is that if the topology changes more frequently than the time needed to reach the steady-state (each change of the true $A_\star$ ``restarts" the algorithm), then a fairly large step-size $\mu$ would be needed. We illustrate this behavior in Fig.~\ref{fig:perturbe1}. The ability of the algorithm to adapt to the true state $\theta^\star$ is also important for its performance due to the fact that we have performance guarantees when $\blambda_i$ reaches its steady-state. The adaptation ability is also evident from the social learning update itself~(\ref{eq:adapt_adaptive})--(\ref{eq:combine}). In~\cite{bordignon2020adaptive}, the authors discuss that small $\delta$ leads to an increased confidence of beliefs, but it comes at the cost of an increased adaptation time, on the order of $\approx {\log 2} / \delta$. Thus, as long as the environment does not change its state faster than the adaptation period, the  performance of the algorithm stays close to what is predicted under Theorems~\ref{thm:conv},~\ref{thm:conv2}. We illustrate this scenario in Fig.~\ref{fig:perturbe2}. 

\section{Application to Twitter Data}
\label{sec:twitter}
    In this section, we apply the graph social learning algorithm introduced in this manuscript to actual datasets. We choose Twitter as a suitable social media platform where we can analyze the posts of users and observe their interactions. In particular, we aim to detect the centrality (i.e., Perron entry) and influence of agents over a network, and identify the most influential agents. We choose this objective as a more realistic goal for Twitter data, rather than trying to recover the entire combination matrix. This is primarily due to the fact that we can only have access to the true adjacency matrix of the users on Twitter, that is to say, we only know who follows whom, but we do not know the weights (confidence levels) agents assign to each other. 
    In the literature, there have been multiple studies trying to analyze the influence of Twitter accounts on various topics. The work \cite{polisci_twitter} analyzed the propagation of information through Twitter networks, and how different users take part in the effective dissemination of ideas. Moreover, the work \cite{twitter_quercia2011mood} analyzes the influential users on Twitter by investigating the linguistic aspects of the tweets of users, such as the grammatical structure and vocabulary of the tweets. However, these studies lack a mathematical foundation to analyze the ``influence" of agents, and rely largely on heuristics. In this context, the work \cite{twitter_influence_measure} aims to leverage ten different attributes of Twitter accounts and their tweets to develop a Twitter-based influence measure. Similarly, the work \cite{twitter_influence_swedish} uses descriptive statistics of the Twitter accounts such as their follower counts, frequency of their posts and the total number of replies, likes and retweets of users. However, what we propose in this study is significantly different. In the proposed algorithm, we do not need to have access to any of these features and the various statistics. In fact, the only input the proposed algorithm needs is the publicly shared tweets of the users. This information is sufficient to learn the centrality of users over a subnetwork of Twitter users, and to identify the most influential user for the formed opinions, using the described mathematical model of social learning.
    
    To identify the influence of agents in a network, we first need to (i) create a subnetwork of Twitter users that is strongly connected, (ii) obtain the tweets (posts) of the users in the created network, and (iii) process the text in the tweets to obtain the log-belief ratios $\boldsymbol{\Lambda}_i$ over time. After these pre-processing operations, we run the proposed algorithm, and estimate the underlying combination matrix of the network of users and the likelihood models of these users. Then, we compute the Perron eigenvector of the estimated combination matrix. We refer to this vector as the ``learned" Perron vector, not to be confused with the ``original" Perron vector of the true combination matrix of the network of users. Note that, there is no ground truth regarding the confidence agents place on each other, hence one cannot possibly know the true combination matrix. Therefore, we heuristically form a combination matrix to obtain its Perron vector, through the procedure of Section \ref{sec:twitter}-\ref{sec:network_formation}. However, as an indisputable ground truth data for a given subnetwork, we know the agents that have the highest number of followers within that subnetwork. In our experiments, these agents happen to coincide with the highest original Perron vector entries.
    
\begin{figure*}
    \begin{subfigure}{.33\textwidth}
      \centering
      \includegraphics[width=\linewidth]{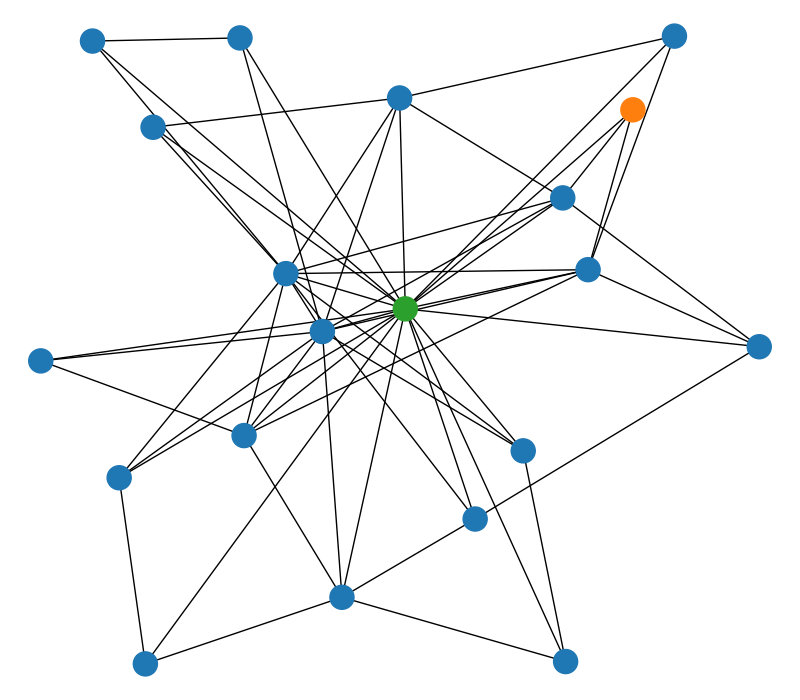}
      \caption{Network for Elon Musk.}
      \label{fig:network_sfig1}
    \end{subfigure}
    \begin{subfigure}{.33\textwidth}
      \centering
      \includegraphics[width=\linewidth]{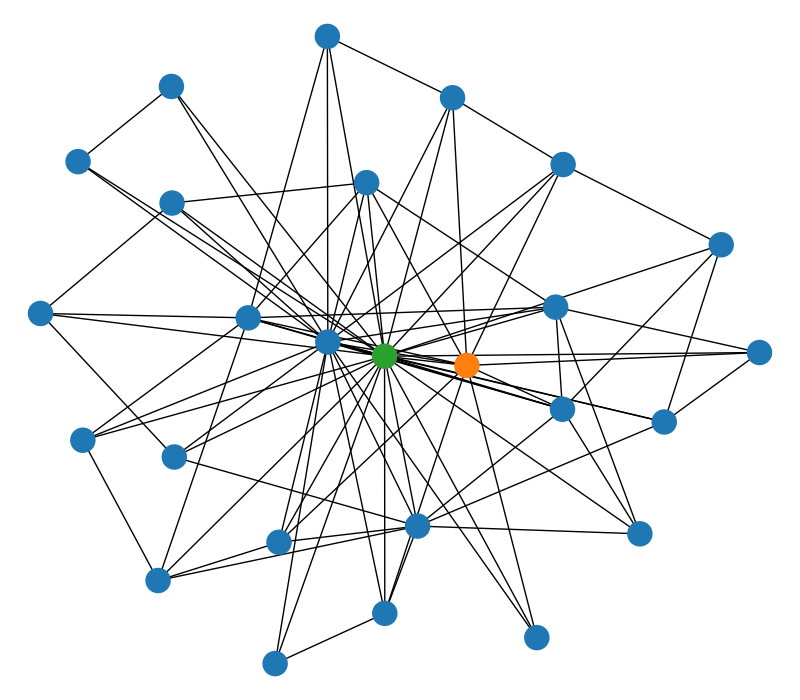}
      \caption{Network for Maggie Haberman.}
      \label{fig:network_sfig2}
    \end{subfigure}
    \begin{subfigure}{.33\textwidth}
      \centering
      \includegraphics[width=\linewidth]{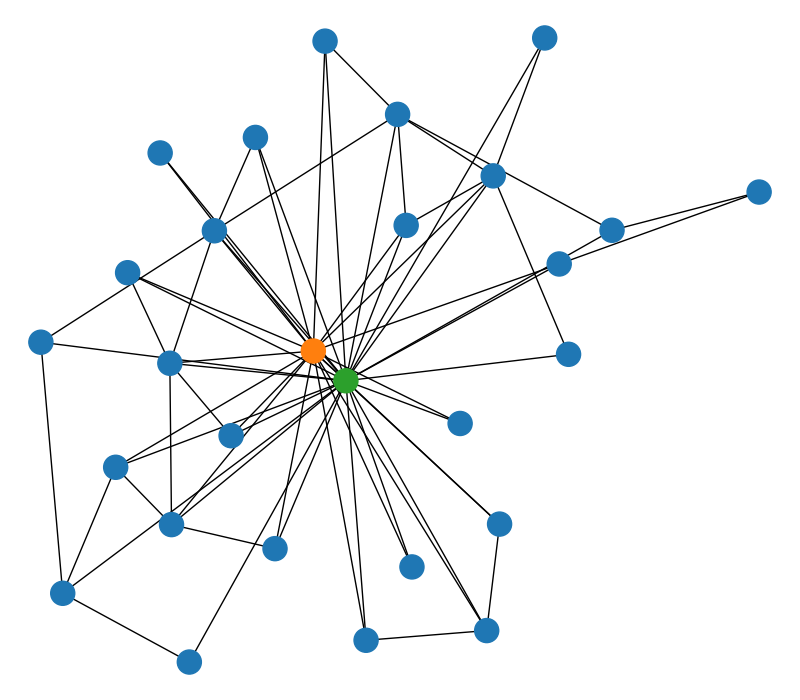}
      \caption{Network for Ben Shapiro.}
      \label{fig:network_sfig3}
    \end{subfigure}
    \captionsetup{justification=justified}
    \caption{Network of agents. Agent $A$ is indicated with color green, and agent $B$ is indicated in color orange according to the description in Section \ref{sec:network_formation}.}
    \label{fig:twitter_network_figures}
    \end{figure*}

    We run experiments on three different user subnetworks of Twitter users by using Twitter API, where we utilize the Tweepy library to build up our queries in Python.
    Quantitatively, we show that we can identify the most central user in all three networks, i.e., the largest entries of the learned Perron vector and the original Perron vector. Note that in these experiments with real data, we do not have access to the likelihood models of the users, hence we do not have ``ground truths" for the influence of agents, i.e., $u_k D_{\textrm{KL}}(L_k(\theta^\star)||L_k(\theta))$. However, to obtain a qualitative measurement, we construct the aforementioned three networks (see Fig.~\ref{fig:twitter_network_figures}) around famous public figures (the first network is built around Elon Musk (CEO, 118.5M followers), the second network is built around Maggie Haberman (journalist, 1.6M followers), and the third network is built around Ben Shapiro (columnist, 5M followers)).
    In all of these networks, we recover these public figures as the most influential users.
    
    \subsection{Network formation}
    \label{sec:network_formation}
        To construct a network, we first select a popular Twitter account $A$, such as an influential CEO or a journalist whom we expect to be influential among other users. Next, in order to {select} a strongly connected network, we build a network starting from a less centered Twitter account $B$ that is followed by that popular user $A$. Starting from account $B$, we construct a subnetwork of depth 2, i.e. we identify 1K followers of account $B$, and 1K followers of each of those followers. While satisfying these conditions, we {filter} users who post frequently, so that they provide sufficient data, and who are less centered (they have less than 10K followers). Among all of these identified follower-following relationships, we construct a network, and verify that the network is strongly connected. Following this procedure, the networks constructed for Elon Musk, Maggie Haberman and Ben Shapiro are of sizes 20, 26 and 28, respectively.

        After constructing the network, we obtain its adjacency matrix by finding out who follows whom. Since there is no way in real world experiments to determine the confidence agents assign to each other, there is no ground truth for the combination matrix. Therefore, we assume that agents assign uniform weights to their neighbors (i.e., to the people they follow). This corresponds to applying the averaging rule \cite[Chapter 14]{Sayed_2014} to the adjacency matrix. 
        
    \subsection{Obtaining user posts}
    
        In order to obtain the tweets of the users, we first build up our query for the Twitter API. This query includes a specific keyword for each network so that we do not fetch unrelated tweets. For the network containing Elon Musk, we choose the keyword to be ``coin OR bitcoin OR crypto-currency", for the tweets between 01.01.2017 and 01.05.2022. In this case, we would like to see the influential agents in that network in shaping the opinion of Twitter users on crypto-currency related matters. For the other networks, we choose the keyword to be ``Trump" for the tweets between 01.01.2017 and 31.01.2021, and ``Biden" for the tweets between 01.01.2021 and 01.05.2022. Hence, we aim to determine the influence of agents in determining the attitude of their respective networks regarding the issue of ``the current president of the United States".
    
    \subsection{Obtaining the log-ratio beliefs of users}
    
        After obtaining the tweets of the users, we need to process the text in those tweets to obtain the intermediate belief vectors. For this purpose, we use a language model based on Roberta, which is trained with around 124M tweets, and fine tuned with the TweetEval benchmark for the sentiment analysis task \cite{roberta_language_model}. We can feed a text to this model and obtain three different probabilities for three different sentiments of the input text: $p_0$ for ``Negative", $p_1$ for ``Neutral", and $p_2$ for ``Positive". 
        We eliminate the neutral labels and recalculate the probability of the positive sentiment of a text as  \begin{equation}
            p\triangleq\frac{p_2}{p_0+p_2}
        \end{equation}
        and accordingly, the probability of the negative sentiment as
        \begin{equation}
            1-p =\frac{p_0}{p_0+p_2}
        \end{equation}
        In Table \ref{table:tweet_sentiments}, we show sample tweets from the network related to Bitcoins. When $p$ is close to 1, this means that the text asserts a positive attitude towards Bitcoin, and when $p$ is closer to 0, this means that the text asserts a negative attitude.

        To construct the log-belief ratio of an agent at a day $i$, we incorporate the sentiment of all tweets within this day. We denote the number of tweets user $k$ shares at iteration $i$ by $N_{k,i}$. 
        Among those tweets, we denote the positive sentiment probability of each tweet $t$ by $p_{k,i,t}$, so that $p_{k,i,t}$ approximates belief $\boldsymbol{\psi}_{k,i,t}(\theta_0)$, and $1-p_{k,i,t}$ as approximates belief  $\boldsymbol{\psi}_{k,i,t}(\theta_1)$. Here, $\theta_0$ is the hypothesis that the underlying topic (e.g., Bitcoin, the current president of the United States, etc.) is ``good'', and $\theta_1$ is the counter-hypothesis. Then, we construct log-belief of each agent $k$ as:
        \begin{equation}
            \boldsymbol{\Lambda}_{k,i} = \frac{1}{N_{k,i}}\sum_{t=1}^{N_{k,i}}\log\left(\frac{p_{k,i,t}}{1-p_{k,i,t}}\right)
        \label{eqn_beliefs}
        \end{equation}
        Note that if $N_{k,i} = 0$ for some agent $k$ at some iteration $i$, we set $\boldsymbol{\Lambda}_{k,i}$ to previous value of it, i.e. $\boldsymbol{\Lambda}_{k,i} = \boldsymbol{\Lambda}_{k,i-1}$.

        \begin{table}
            \begin{tabularx}{\linewidth}{@{} X C @{}}
                \toprule
                Sample Tweets & Positive Sentiment \\
                \midrule
                \emph{\#Bitcoin for the win} & $p = 0.99$ \\
                \addlinespace
                \emph{Unless otherwise stated, I am adding Bitcoin every week} & $p = 0.82$ \\
                \addlinespace
                \emph{Bitcoin falls from \$66K highs, Tesla down 3\% after Elon Musk warns he could sell more \#Bitcoin} & $p=0.02$ \\
                \bottomrule
            \end{tabularx}
            \caption{Examples of sentiment analysis and generated sentiment probabilities.}
            \label{table:tweet_sentiments}
        \end{table}

        \begin{figure*}[h]
            \begin{subfigure}{.33\textwidth}
              \centering
              \includegraphics[width=\linewidth]{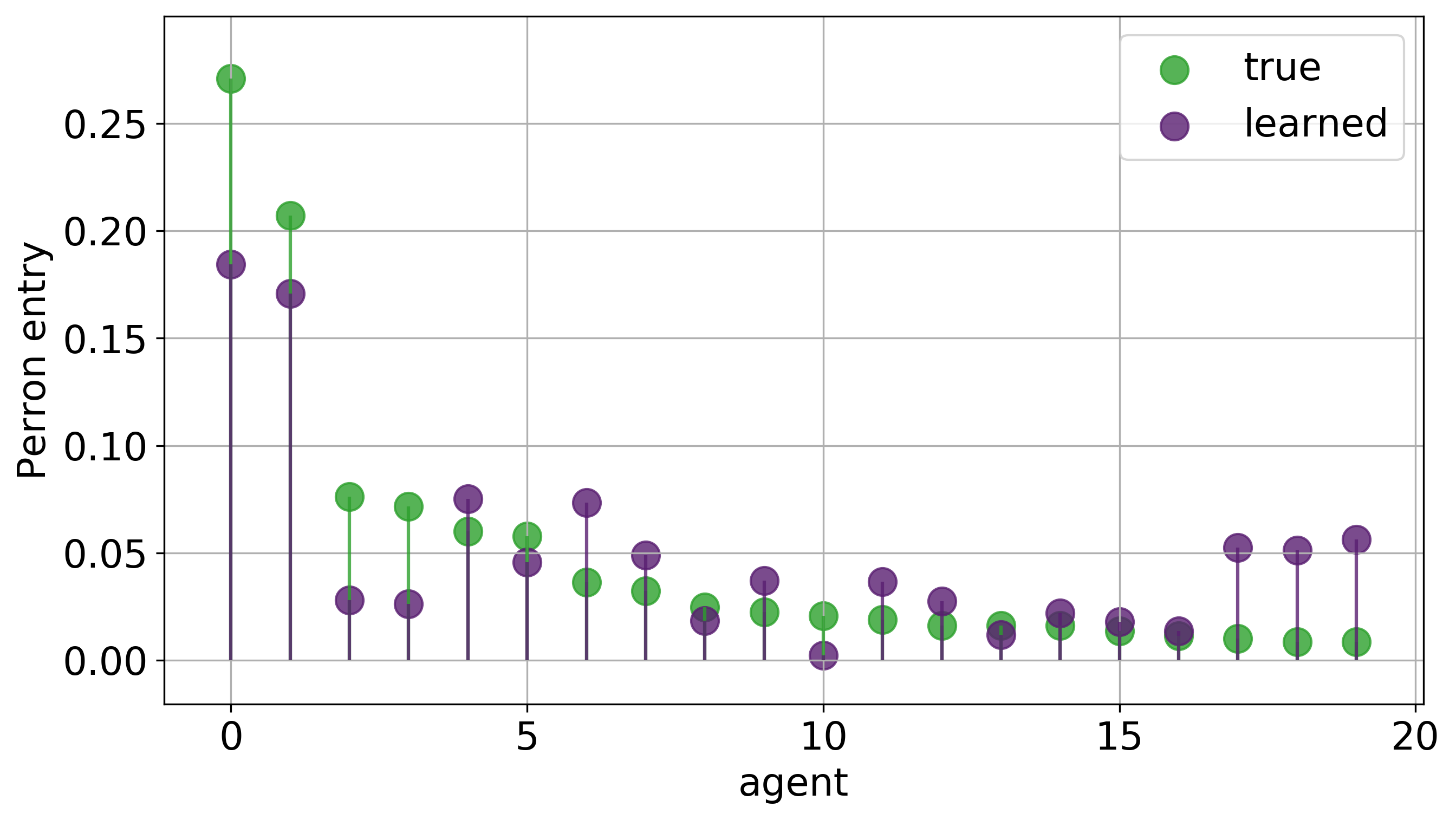}
              \caption{Network of Elon Musk (agent 0).}
              \label{fig:sfig1}
            \end{subfigure}
            \begin{subfigure}{.33\textwidth}
              \centering=
              \includegraphics[width=\linewidth]{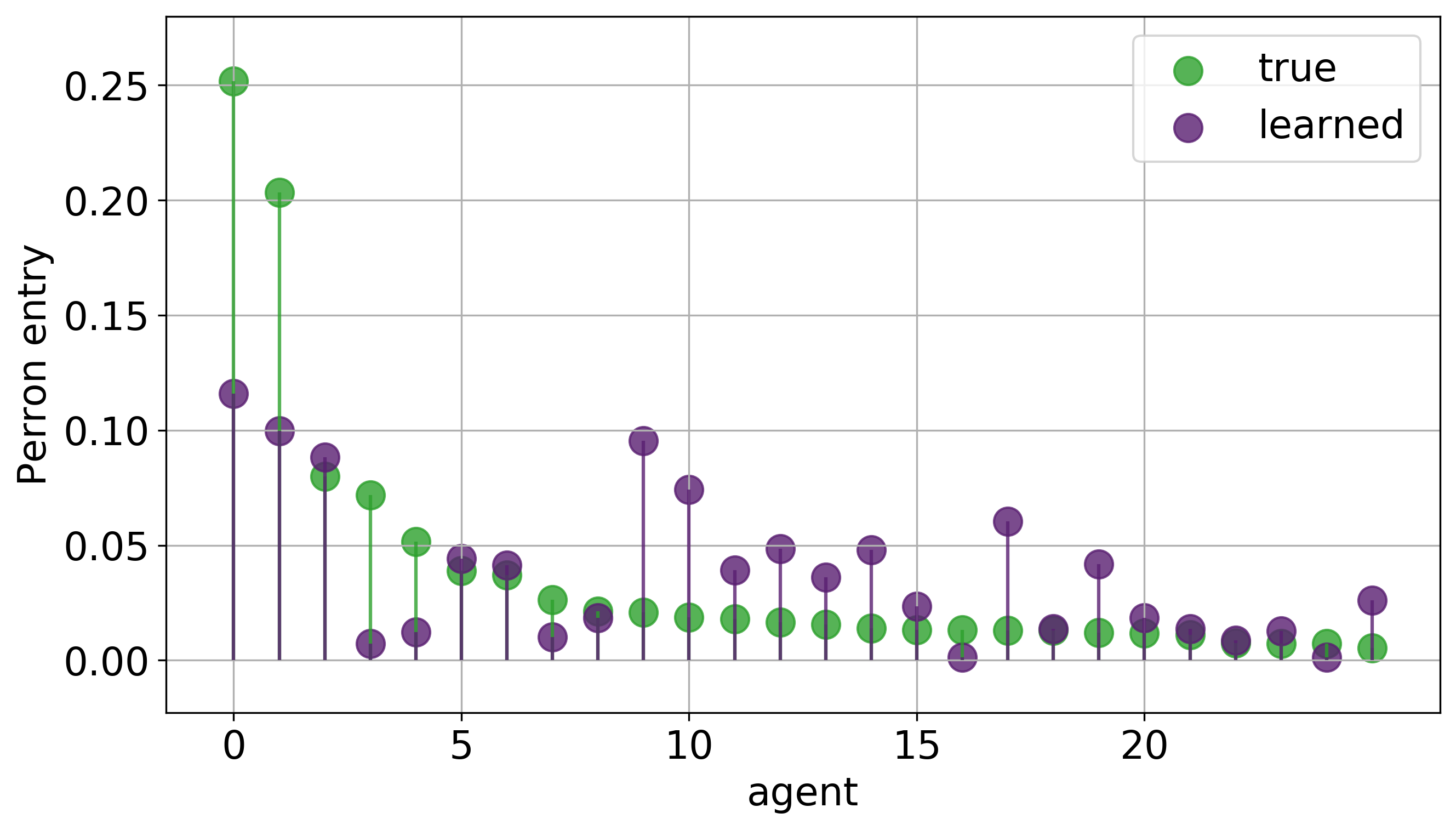}
              \caption{Network of Maggie Haberman (agent 0).}
              \label{fig:sfig2}
            \end{subfigure}
            \begin{subfigure}{.33\textwidth}
              \centering
              \includegraphics[width=\linewidth]{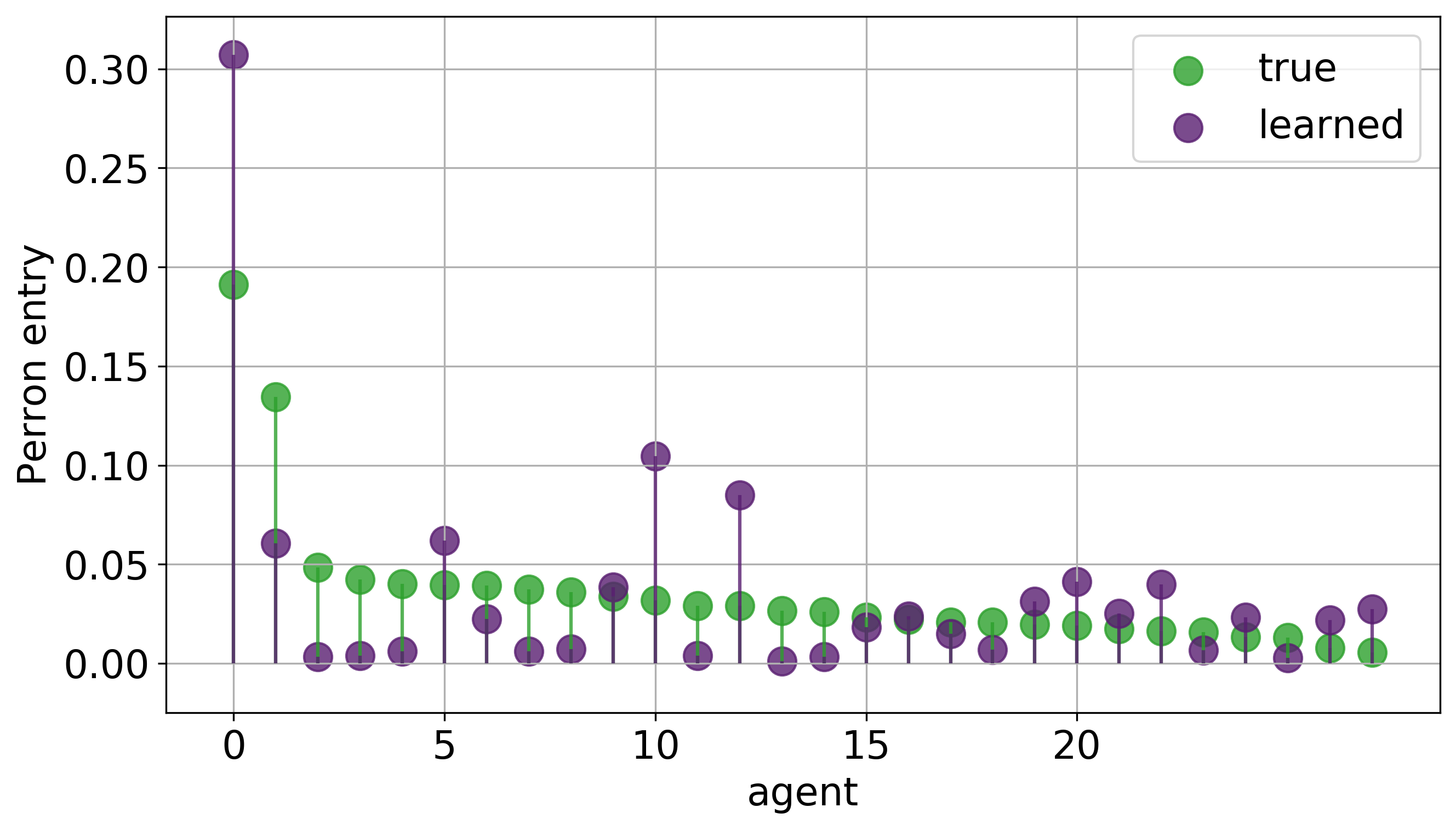}
              \caption{Network of Ben Shapiro (agent 0).}
              \label{fig:sfig3}
            \end{subfigure}
            \captionsetup{justification=justified}
            \caption{Plots of agent centralities. Green points are the entries of the original Perron vector, and purple points belong to the learned Perron vectors. In all networks, we see that the algorithm identifies the most central agent in the graph.}
            \label{fig:twitter_centrality_figures}
        \end{figure*}
        
        \begin{figure*}[h]
            \begin{subfigure}{.33\textwidth}
              \centering
              \includegraphics[width=\linewidth]{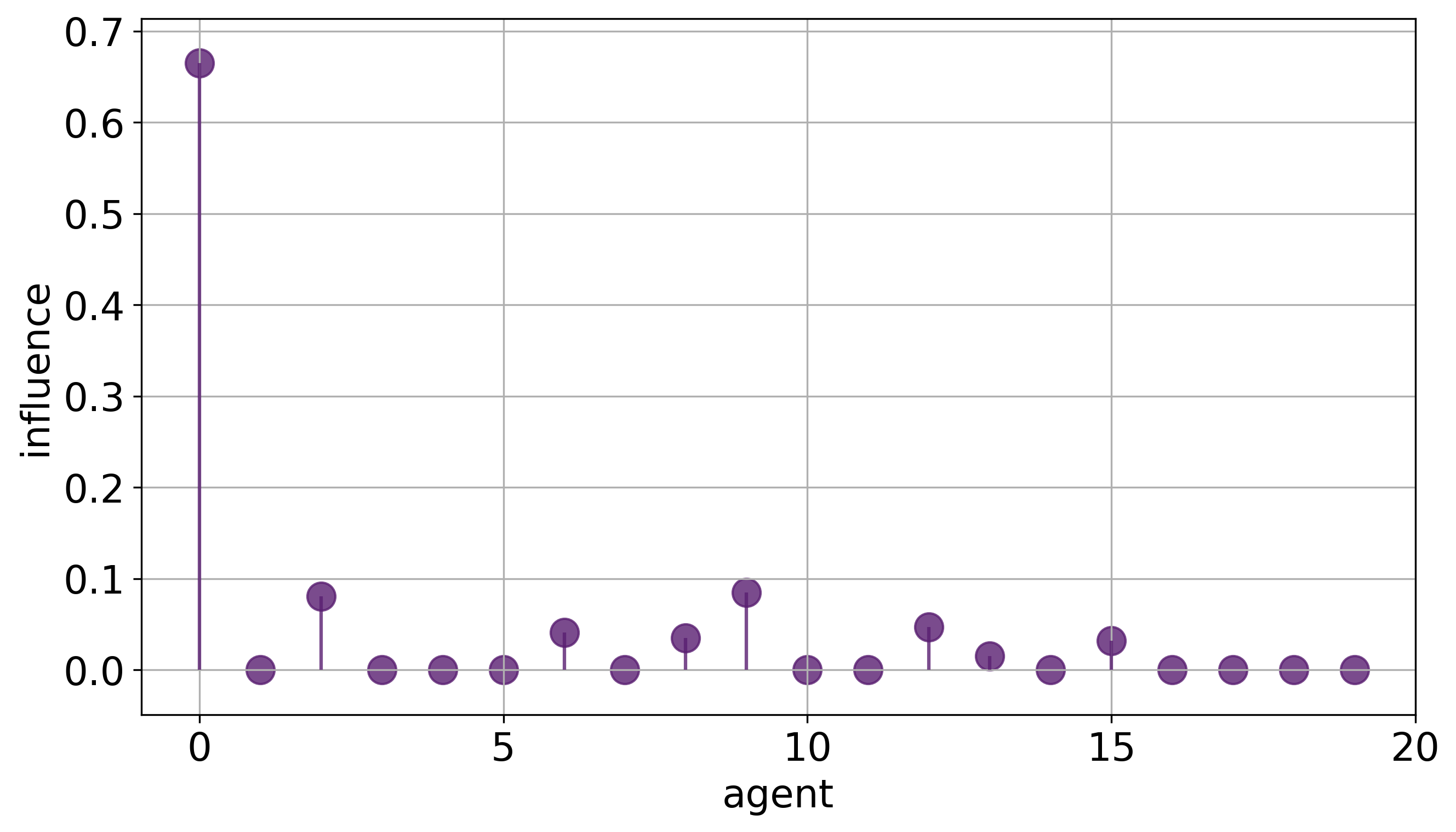}
              \caption{Network of Elon Musk (agent 0).}
              \label{fig:influence_sfig1}
            \end{subfigure}
            \begin{subfigure}{.33\textwidth}
              \centering
              \includegraphics[width=\linewidth]{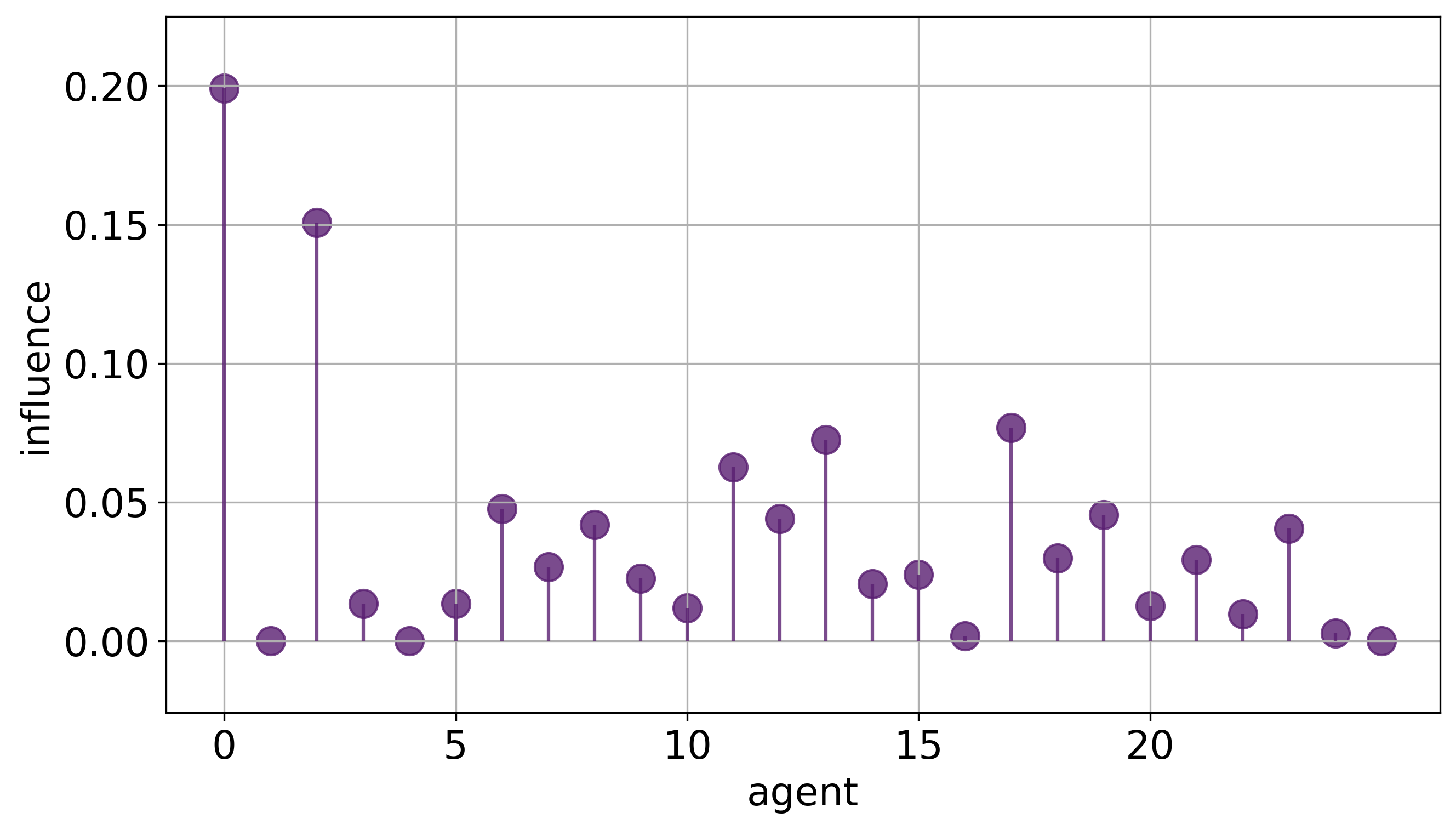}
              \caption{Network of Maggie Haberman (agent 0).}
              \label{fig:influence_sfig2}
            \end{subfigure}
            \begin{subfigure}{.33\textwidth}
              \centering
              \includegraphics[width=\linewidth]{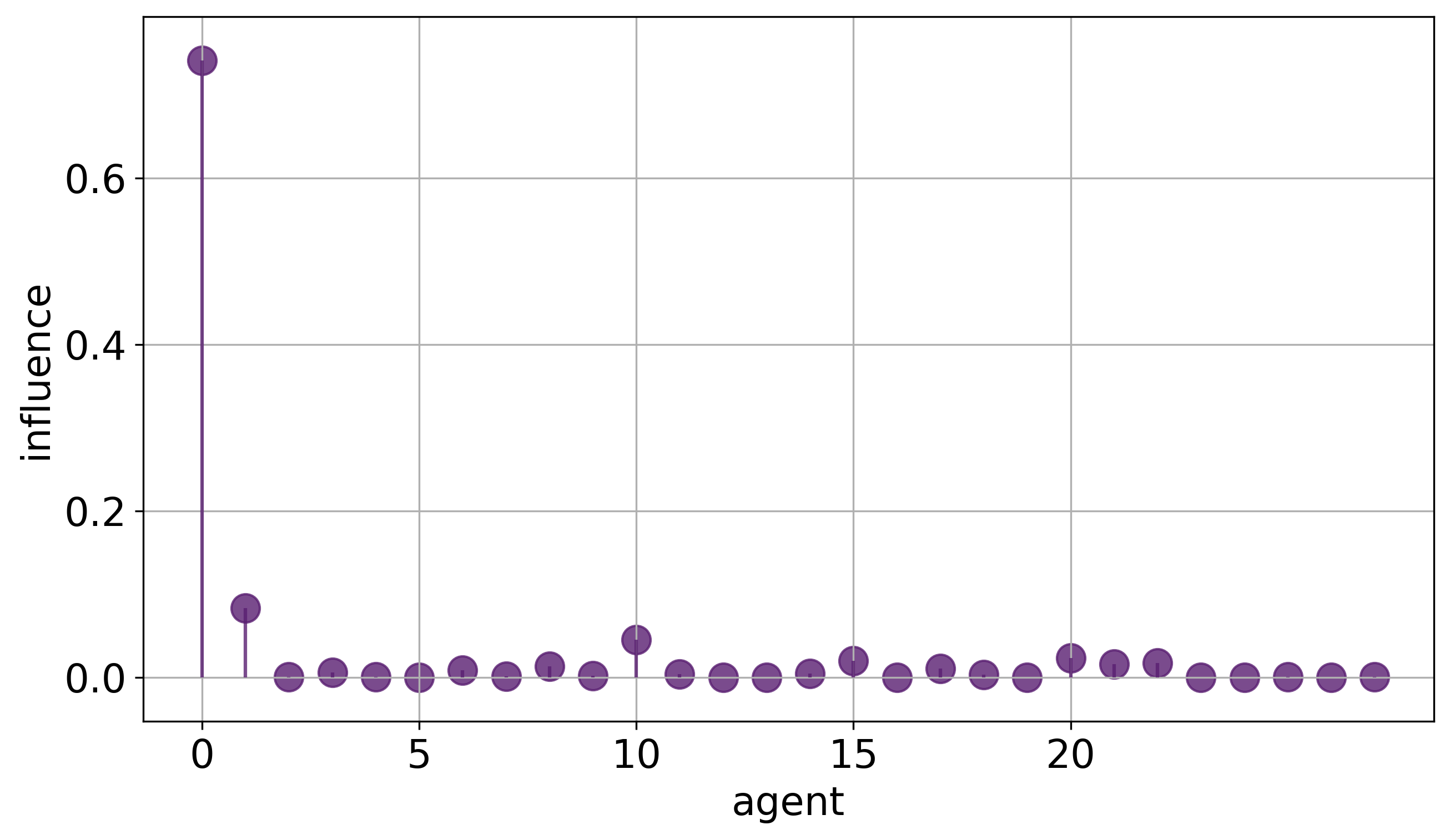}
              \caption{Network of Ben Shapiro (agent 0).}
              \label{fig:influence_sfig3}
            \end{subfigure}
            \captionsetup{justification=justified}
            \caption{Plots of agent influences. In the respective networks containing Elon Musk (CEO -- 110M followers on Twitter), Maggie Haberman (journalist -- 1.7M followers on Twitter) and Ben Shapiro (columnist -- 4.5M followers on Twitter), we can identify these public figures as the most influential agents. For instance, in Fig. \ref{fig:influence_sfig1}, we can identify Elon Musk as the most influential Twitter user in the corresponding subnetwork}.
            \label{fig:twitter_influence_figures}
        \end{figure*}
    
    \subsection{Adjustments to the algorithm}
    
        We have made some adjustments to the graph social learning algorithm to cope with real life data. In particular, instead of performing gradient updates at each iteration, as in the online stochastic learning algorithm, we use stochastic mini-batches. Namely, we select a window size $W$, and average the gradients calculated within that window, and then perform the gradient step. In our experiments, we observe that this practice reduces the noise in the gradients due to real life data. Furthermore, we use $\ell_1$ regularization to promote sparsity. The motivation for sparsity is to get rid of unnecessary links between different users in the graph. 
        These hyperparameters in the experiments are determined with a grid search on one of the networks (Ben Shapiro), and then the parameters are used in the other two networks. These hyperparameters are: for the stochastic mini-batch size, $W=30$, for the stepsize $\delta=0.0001$, for the learning rate $\mu = 0.0003$ and for the $\ell_1$ regularization weight $\alpha=0.006$. The fact that we obtain desirable outcomes in all cases suggests that the algorithm  generalizes and performs well across different networks.
    
    \subsection{Experimental results}
    
        We compare the Perron vectors of the learned combination matrix and the original combination matrix (which is constructed by assigning uniform weights to users each user follows). In these  comparisons of Perron vectors, we show that we can correctly estimate the most central agents in all three networks. These comparisons can be seen in Fig. \ref{fig:twitter_centrality_figures}. Secondly, we show the influence plots of the agents in these three networks in Fig. \ref{fig:twitter_influence_figures}. These figures, ``qualitatively" show that we can indeed identify the popular accounts (Elon Musk, Maggie Haberman and Ben Shapiro) as the most influential agents in their respective networks. For instance, in Fig. \ref{fig:influence_sfig1}, we can identify Elon Musk as the most influential Twitter user of his network of users.

\section{Conclusions}
    In this study, we show that a sequence of publicly exchanged beliefs in the adaptive social learning protocol contains rich information about the underlying model. We present an algorithm for learning the agents' informativeness in terms of KL-divergences between likelihood models, and for identifying a combination graph. We demonstrate that these quantities determine the probability of error of the true hypothesis estimator, and we introduce a notion of a global agent influence, which quantifies the individuals' contribution to learning. As a result, the suggested approach enables us to determine the most influential agents in the opinion formation process. We also describe how to apply the algorithm to Twitter data. Our experiments on both synthetic data and Twitter data illustrate that we can accurately find global influencers and learn the underlying graph. 
    
\section*{Acknowledgement}
    We thank our colleague Virginia Bordignon for her valuable suggestions regarding the application to Twitter data.
\appendices

\section{Proof of Lemma~\ref{lemma:mean}}\label{apx:mean}
    From Markov's inequality, we have
    \begin{align}
        &\;\mathbb P\left(\Bigg\| \frac 1M \sum_{j=i-M}^{i-1} \blambda_{j} - \bE \blambda \Bigg\|_{\textrm F}^2 > \varepsilon\right) \nonumber\\
        &\; \leq \frac 1{\varepsilon} \bE \Bigg\| \frac 1M \sum_{j=i-M}^{i-1} \blambda_{j} - \bE \blambda \Bigg\|_{\textrm F}^2 \nonumber\\
        &\; = \frac{1}{M^2\varepsilon} \bE \Bigg\| \sum_{j=i-M}^{i-1} \Bigg(\blambda_{j} - \bE \blambda \Bigg)\Bigg\|_{\textrm F}^2.
        \label{eq:l1_ineq}
    \end{align}
    The expectation above can be expanded into:
    \begin{align}
        &\;\bE \Bigg\| \sum_{j=i-M}^{i-1} \Bigg(\blambda_{j} - \bE \blambda \Bigg) \Bigg\|_{\textrm F}^2 = \sum_{j=i-M}^{i-1} \bE \Big\| \blambda_{j} - \bE \blambda \Big\|_{\textrm F}^2\nonumber\\
        &\; + 2 \sum_{\substack{j_1, j_2 = i - M,\\ j_1 < j_2}}^{i-1} \textrm{ Tr} \left( \bE \left(\blambda_{j_1} - \bE \blambda \right) \left(\blambda_{j_2} - \bE \blambda \right)^\bT \right).
    \end{align}
    Introducing the history 
    \begin{align}
        \bF_{i} \triangleq \{\boldsymbol\zeta_{k, j},\; j < i,\; \forall k\in \mathcal N\},
        \label{eq:history}
    \end{align}which collects all observations up to time $i$, and conditioning over $\bF_{j_1+1}$, we have 
    \begin{align}
        &\;\bE \left(\blambda_{j_1} - \bE \blambda \right) \left(\blambda_{j_2} - \bE \blambda \right)^\bT \nonumber\\
        =&\; \bE \left(\left(\blambda_{j_1} - \bE \blambda \right) \bE \left(\blambda_{j_2}^\bT - \bE \blambda^\bT \big| \bF_{j_1 + 1} \right)\right)
        \label{eq:l1_h}
    \end{align}
    This equality holds since $\bE \left(\blambda_{j_1} | \bF_{j_1 + 1}\right) = \blambda_{j_1}$.
    Using the main recursion formula~(\ref{eq:recursion_adaptive}), we can represent $\blambda_{j_2}$ in terms of $\blambda_{j_1}$ and observations $\bL_{t}$ with $j_1 + 1 \leq t\leq j_2$:
    \begin{align}
        \blambda_{j_2} =&\; (1-\delta)^{j_2-j_1} (A_\star^{j_2-j_1})^\bT \blambda_{j_1} \nonumber\\
        &\; + \delta \sum_{t=0}^{j_2-j_1-1} (1-\delta)^t (A_\star^t)^\bT  \bL_{j_2-t}
    \end{align}
    It follows that
    \begin{align}
        \bE \left(\blambda_{j_2}\big| \bF_{j_1 + 1}\right) =&\; (1-\delta)^{j_2-j_1} (A_\star^{j_2-j_1})^\bT \blambda_{j_1} \nonumber\\
        &\; + \delta \sum_{t=0}^{j_2-j_1-1} (1-\delta)^t (A_\star^t)^\bT \bar \cL
        \label{l2:h11}
    \end{align}
    where $\bar \cL$ is the expected value of $\bL_i$ defined in~(\ref{eq:barL}).
    Using~(\ref{l2:h11}), we can rewrite~(\ref{eq:l1_h}) as:
    \begin{align}
        &\;\bE \left(\left(\blambda_{j_1} - \bE \blambda \right) \bE \left(\blambda_{j_2}^\bT - \bE \blambda^\bT \big| \bF_{j_1 + 1} \right)\right) \nonumber\\
        =&\; \bE \left(\blambda_{j_1} - \bE \blambda \right) \left((1-\delta)^{j_2-j_1} (A_\star^{j_2-j_1})^\bT \blambda_{j_1} - \bE \blambda \right)^\bT \nonumber\\
        &\; + \bE \left(\blambda_{j_1} - \bE \blambda \right) \left( \delta \sum_{t=0}^{j_2-j_1-1} (1-\delta)^t (A_\star^t)^\bT \bar \cL \right)^\bT \nonumber\\
        =&\; \bE \left(\blambda_{j_1} - \bE \blambda \right) \left( \blambda_{j_1} - \bE \blambda \right)^\bT (1-\delta)^{j_2-j_1} A_\star^{j_2-j_1} \nonumber\\
        &\; + \bE \left(\blambda_{j_1} - \bE \blambda \right) \bE\blambda^\bT \left((1-\delta)^{j_2-j_1} A_\star^{j_2-j_1} - I\right) \nonumber\\
        &\; + \bE \left(\blambda_{j_1} - \bE \blambda \right) \left( \delta \sum_{t=0}^{j_2-j_1-1} (1-\delta)^t (A_\star^t)^\bT \bar \cL \right)^\bT \nonumber\\
        =&\;\bE \left(\blambda_{j_1} - \bE \blambda \right) \left( \blambda_{j_1} - \bE \blambda \right)^\bT (1-\delta)^{j_2-j_1} A_\star^{j_2-j_1}
        \label{eq:l1_trace}
    \end{align}
    where the last equation holds since in steady state $\bE \blambda_{j_1} = \bE \blambda$. The trace of~(\ref{eq:l1_trace}) then becomes\footnotemark:
    \begin{align}
        &\;\textrm{Tr } \bE \left(\blambda_{j_1} - \bE \blambda \right) \left( \blambda_{j_1} - \bE \blambda \right)^\bT (1-\delta)^{j_2-j_1} A_\star^{j_2-j_1}\nonumber\\
        &\;\leq \bE \Big(\big\|\blambda_{j_1} - \bE \blambda \big\|_{\textrm F} \nonumber\\
        &\;\;\;\;\; \times \big\| (1-\delta)^{j_2-j_1} (A_\star^{j_2-j_1}) \left(\blambda_{j_1} - \bE \blambda \right)\big\|_{\textrm F} \Big) \nonumber\\
        &\;\leq (1-\delta)^{j_2 - j_1} \|A_\star^{j_2-j_1}\|_{\textrm F} \bE \|\blambda_{j_1} - \bE \blambda\big\|^2_{\textrm F} \nonumber\\
        &\;\leq (1-\delta)^{j_2-j_1} \sqrt{|\mathcal N|} \bE \big\| \blambda_{j_1} - \bE \blambda\big\|^2_{\textrm F},
    \end{align}
    where the last inequality holds because $A_\star$ and its powers are left-stochastic, i.e. all the entries are non-negative and each column's entries add up to one:
    \begin{align}
        \|A_\star\|_{\textrm F}^2 = \sum_k \sum_\ell a_{k\ell}^2 \leq \sum_k \left(\sum_\ell a_{k\ell}\right)^2 = \sum_k 1 = |\mathcal N|
    \end{align}
    Next, let us study the following sum:
    \begin{align}
        &\;\sum_{\substack{j_1, j_2 = i - M,\\ j_1 < j_2}}^{i-1} (1-\delta)^{j_2-j_1} \nonumber\\
        &\;= (M-1)(1-\delta) + (M-2)(1-\delta)^2 + \dots + \nonumber\\
        &\;\;\;\;\; + 2(1-\delta)^{M-2} + (1-\delta)^{M-1} \nonumber\\
        &\;= \frac{M-1 - (1-\delta)^M}{\delta} - (1-\delta)\delta^{M-2} = O(M/\delta)
    \end{align}
    Additionally, under the steady-state condition~(\ref{eq:blambda_lim_}), the following property for $\blambda$ holds:\footnotetext{Using $\textrm{Tr}\left(AB^\bT\right) \leq \|A\|_{\textrm F} \|B\|_{\textrm F}$.}
    \begin{align}
        &\;\bE \| \blambda- \bE \blambda \|^2_{\textrm F} \nonumber\\
        &\;= \bE\; \Tr \left(\left(\blambda- \bE \blambda\right)\left(\blambda- \bE \blambda\right)^\bT\right) \nonumber\\
        &\;\overset{(\ref{eq:blambda_lim_})}{=}
        \bE \; \Tr \Big( \delta \sum_{t_1=0}^\infty \left(1-\delta\right)^{t_1}(A_\star^{t_1})^\bT\left(\bL_{t_1} - \bar \cL\right)\nonumber\\
        &\;\;\;\;\; \times  \delta \sum_{t_2=0}^\infty \left(1-\delta\right)^{t_2}(A_\star^{t_2})^\bT\left(\bL_{t_2} - \bar \cL \right) \Big) \nonumber\\
        &\;= \delta^2 \bE \; \Tr \Big( \sum_{t_1=0}^\infty \sum_{t_2=0}^\infty (1-\delta)^{t_1 + t_2} (A_\star^{t_1})^\bT (\bL_{t_1} -\bar\cL) \nonumber\\
        &\;\;\;\;\;\times (\bL_{t_2} -\bar\cL)^\bT A_\star^{t_2} \Big) \nonumber\\
        &\;= \delta^2 \sum_{t=0}^\infty (1-\delta)^{2t} \Tr\left((A_\star^{t})^\bT \R_{\bL} A_\star^t\right) = O(\delta)
        \label{eq:Ellll}
    \end{align}
    due to i.i.d. $\bL_i$.
    Combining the derivations above, expression~(\ref{eq:l1_ineq}) becomes:
    \begin{align}
        &\;\mathbb P\left(\Bigg\| \frac 1M \sum_{j=i-M}^{i-1} \blambda_{j} - \bE \blambda \Bigg\|_{\textrm F}^2 > \varepsilon\right) \nonumber\\
        &\;\leq \frac{1}{M^2\varepsilon} \sum_{j=i-M}^{i-1} \bE \big\| \blambda_{j} - \bE \blambda \big\|_{\textrm F}^2\nonumber\\
        &\; + \frac{2}{M^2\varepsilon} \sum_{j_1 < j_2} (1-\delta)^{j_2-j_1}  \sqrt{|\mathcal N|} \bE \big\| \blambda_{j_1} - \bE \blambda\big\|^2_{\textrm F}\nonumber\\
        &\;= \frac{1}{M \varepsilon}\left(1 +  \frac{2 \sqrt{|\mathcal N|}}{M} \sum_{j_1 < j_2} (1-\delta)^{j_2-j_1} \right)\bE \| \blambda- \bE \blambda \|^2_{\textrm F} \nonumber\\
        &\;= \frac{1}{M \varepsilon}\left(1 +  O(1/\delta) \right)O(\delta) \nonumber\\
        &\; = O\left(\frac 1 {M\varepsilon}\right) 
        \label{eq:dev1111}
    \end{align}
    since $ \sum_{j_1 < j_2} (1-\delta)^{j_2-j_1} = O(M/\delta)$.
    Taking $M\rightarrow\infty$, the right-hand side of~(\ref{eq:dev1111}) is zero for any $\varepsilon > 0$. Thus, we establish convergence in probability by definition.

\section{Proof of Lemma~\ref{lemma:risk}}\label{apx:risk}
    Under some reasonable assumptions on the distribution of the random variables, as required by the dominated convergence theorem in mathematical analysis, it is possible to exchange the expectation and gradient operations~\cite{sayed_2023}. Thus, using~(\ref{eq:cost_function}), the gradient of~(\ref{eq:risk}) is given by:
    \begin{align}
        \nabla J_i(A)
        =& -(1-\delta) \bE \left[\bdelta_{i-1} \left( \bdelta_i^\bT -(1-\delta) \bdelta_{i-1}^\bT A \right)\right]
        \label{eq:grad}
    \end{align}
    We first verify that the risk function $J_i(A)$ has Lipschitz gradients. Note that for any $A_1$, $A_2$:
    \begin{align}
        &\textrm{Tr}\left(\left(\nabla J_i(A_1) - \nabla J_i(A_2)\right)^\bT \left(A_1 - A_2\right)\right)\nonumber\\
        &\;=(1-\delta)^2 \textrm{Tr}\Big(\bE \bdelta_{i-1}\bdelta_{i-1}^\bT \left(A_1 - A_2\right)\left(A_1 - A_2\right)^\bT\Big)\nonumber\\
        &\;\overset{(a)}{\leq} (1-\delta)^2 \lambda_{\max} \left( \bE \bdelta_{i-1}\bdelta_{i-1}^\bT \right) \|A_1 - A_2\|_{\textrm F}^2,
    \end{align}
    where step (a) follows from the following considerations. For matrices $X$ and $Y$ of appropriate dimensions, where $X$ is positive definite with eigendecomposition $X = U_X\Lambda_XU_X^{-1}$, the following inequality holds:
    \begin{align}
        &\;\Tr\left(XYY^\bT\right) = \Tr\left(U_X\Lambda_XU_X^{-1} YY^\bT\right) \nonumber\\
        &\;= \Tr\left(\Lambda_XU_X^{-1} YY^\bT U_X\right) \leq \lambda_{\max}(X) \Tr(U_X^{-1} YY^\bT U_X)\nonumber\\
        &\;= \lambda_{\max}(X) \|Y\|_{\textrm F}^2.
    \end{align}
    Similarly, we can verify that $J_i(A)$ is strongly convex since
    \begin{align}
        &\textrm{Tr}\left(\left(\nabla J_i(A_1) - \nabla J_i(A_2)\right)^\bT \left(A_1 - A_2\right)\right)\nonumber\\
        \geq&\; (1-\delta)^2 \lambda_{\min} \left( \bE \bdelta_{i-1}\bdelta_{i-1}^\bT \right) \|A_1 - A_2\|_{\textrm F}^\bT.
    \end{align}
    \noindent Next we verify that $\bE \bdelta_{i}\bdelta_{i}^\bT$ is positive definite and finite. For this purpose, we refer to~\cite[Lemma 1]{shumovskaia2022explainability}, which establishes that $\blambda$ given by~(\ref{eq:blambda_lim_}) is element-wise bounded as follows:
    \begin{align}
        |\blambda| \preceq \bar \Lambda &\triangleq \delta b \sum_{t=0}^\infty \left(1-\delta\right)^{t}(A_\star^{t})^\bT \mathds{1}\mathds{1}^\bT\nonumber\\
        &= \delta b \left( I - (1-\delta)A_\star^\bT\right)^{-1} \mathds{1}\mathds{1}^\bT
        \label{eq:lambda_bar_}
    \end{align}
    Using~(\ref{eq:lambda_bar_}), we can also bound the sample average:
    \begin{align}
        \Big|\frac 1M \sum_{j=i-M}^{i-1} \blambda_j \Big| \preceq \frac 1M \sum_{j=i_M}^{i-1} |\blambda_j|  = \bar \Lambda
        \label{eq:av_lam}
    \end{align}
    Then, $\bE \bdelta_{i}\bdelta_{i}^\bT$ is also bounded:
    \begin{align}
        &\;\bE \bdelta_i\bdelta_i^\bT \nonumber\\
        &\; = 
        \bE \left(\blambda_i - \frac 1M \sum_{j=i-M}^{i-1} \blambda_j\right) \left(\blambda_i - \frac 1M \sum_{j=i-M}^{i-1} \blambda_j\right)^\bT \nonumber\\
        &\; \preceq \bE \left(|\blambda_i| + \Big|\frac 1M \sum_{j=i-M}^{i-1} \blambda_j\Big|\right)\left(|\blambda_i| + \Big|\frac 1M \sum_{j=i-M}^{i-1} \blambda_j\Big|\right)^\bT \nonumber\\
        &\; \preceq  2 \bar\Lambda \cdot 2 \bar\Lambda^\bT = 4 \bar\Lambda \bar\Lambda^\bT.
    \end{align}
    From Lemma~\ref{lemma:mean} we know that the following convergence in probability holds:
    \begin{align}
        \frac 1M \sum_{i-M}^{i-1}\blambda_j \xrightarrow{M\rightarrow\infty} \bE \blambda
        \label{eq:sum_d}
    \end{align}
    By definition of convergence in probability, for any $\varepsilon > 0$ and $\pi \in (0, 1)$, there exists $M_0$ such that for any $M \geq M_0(\pi)$
    the probability of the event $\omega$:
    \begin{align}
        \omega \triangleq \left\{\Big\|\frac 1M \sum_{i-M}^{i-1}\blambda_j - \bE \blambda\Big\|^2_{\textrm F} \leq \varepsilon\right\}
        \label{eq:omega}
    \end{align}
    is bounded as follows:
    \begin{align}
        \mathbb P (\omega) \geq 1 - \pi.
        \label{eq:eps_ineq}
    \end{align}
    By the law of total expectation and using~(\ref{eq:eps_ineq}), we can represent $\bE \bdelta_i \bdelta_i^\bT$ as:
    \begin{align}
        &\bE \bdelta_i \bdelta_i^\bT \nonumber\\
        &= \bE \left(\bdelta_i \bdelta_i^\bT \big| \omega \right) \mathbb P (\omega) + \bE \left(\bdelta_i \bdelta_i^\bT \big| \bar \omega \right) \mathbb P (\bar \omega)\nonumber\\
        &\geq (1-\pi) \cdot \bE \left(\bdelta_i \bdelta_i^\bT \big| \omega \right) + 0 \cdot \bE \left(\bdelta_i \bdelta_i^\bT \big| \bar \omega \right)
        \label{eq:tobounddd}
    \end{align}
    where $\bar \omega$ denotes the complementary event of $\omega$, and $\bE \left(\bdelta_i \bdelta_i^\bT \big| \bar \omega \right)$ is finite due to finiteness of each $\blambda_i$, as established in~(\ref{eq:lambda_bar_}). Next, using definition~(\ref{eq:bdelta}),
    and the fact that $g(X) = XX^\bT$ is a continuous bounded function given a bounded input$^{\textrm{\footnotemark}}$
    \footnotetext{If a sequence of random variables $\boldsymbol x_1, \boldsymbol
    x_2, \dots$ converges in distribution to a random variable $\boldsymbol x$, then $\bE g(\boldsymbol x_i) \xrightarrow{i\rightarrow\infty} \bE g(\boldsymbol x)$ for a continuous and bounded function $g(\cdot)$}
    we get that:
    \begin{align}
        &\; \bE \left(\bdelta_i \bdelta_i^\bT \big| \omega \right) \nonumber\\
        &\;= \bE \left(\Big(\blambda_i - \frac 1M \sum_{j=i-M}^{i-1} \blambda_j\Big)\Big(\blambda_i - \frac 1M \sum_{j=i-M}^{i-1} \blambda_j\Big)^\bT\Bigg| \omega \right) \nonumber\\
        &\;\overset{(\ref{eq:omega})}{=} \bE \left(\blambda - \bE \blambda + O(\sqrt\varepsilon) \right)\left(\blambda - \bE \blambda + O(\sqrt\varepsilon)\right)^\bT\nonumber\\
        &\;= \bE (\blambda - \bE \blambda)(\blambda - \bE \blambda)^\bT + O(\varepsilon) \nonumber\\
        &\; = \delta^2 \sum_{t=0}^\infty \left(1-\delta\right)^{2t}(A_\star^t)^\bT \R_{\bL} A_\star^{t} + O(\varepsilon) \nonumber\\
        &\; = \delta^2 \R_{\bL} + \sum_{t=1}^\infty \left(1-\delta\right)^{2t}(A_\star^t)^\bT \R_{\bL} A_\star^{t} + O(\varepsilon) \nonumber\\
        &\; \overset{(a)}{\geq} \delta^2 \R_{\bL}  + O(\varepsilon)  \geq \tau\delta^2 I + O(\varepsilon)
        \label{eq:helper222}
    \end{align}
    where the last inequality is due to Assumption~\ref{asm:loglikelihoods_positivedefinite}.
    Thus,~(\ref{eq:tobounddd}) becomes:
    \begin{align}
        \bE \bdelta_i \bdelta_i^\bT \geq (1-\pi) \left(\tau\delta^2 I + O(\varepsilon)\right)
        \label{eq:tobounddd2}
    \end{align}
    Since we can choose $\varepsilon$ as small as possible, let us set $\varepsilon = 1/\sqrt M$.
    Then, using~(\ref{eq:dev1111}), we get that
    \begin{align}
        \mathbb P\left(\bar w\right) 
        =&\; O(1/M\varepsilon) =  O(1/\sqrt M).
        \label{eq:r2}
    \end{align}
    Therefore, we can take the corresponding $\pi$ to be on the order of $1 / \sqrt M$ due to~(\ref{eq:eps_ineq}). Returning to~(\ref{eq:tobounddd2}), we get:
    \begin{align}
        \bE \bdelta_i \bdelta_i^\bT &\;\geq \left(1 + O\left(1/\sqrt M\right) \right)\left(\tau \delta^2 I + O(1/\sqrt M)\right) \nonumber\\
        &\;= \tau \delta^2 I + O\left(1/\sqrt M\right).
        \label{eq:bdeltabdelta_fin}
    \end{align}
    Thus, $\bE \bdelta_i \bdelta_i^\bT$ is positive-definite for $M$ large enought i.e. as long as $M \gg 1/\delta^4 $.
    
    We now examine the relation of the minimizer of $A_{\min,i} \triangleq \min_A J_i(A)$ to the true combination matrix $A_\star$. From strong convexity, the minimizer of $J_i(A)$ is unique and satisfies:
    \begin{align}
        &\;\nabla J_i(A_{\min,i}) = 0
    \end{align}
    Referring to~(\ref{eq:grad}), we get
    \begin{align}
        -(1-\delta) \bE \bdelta_{i-1} \left(\bdelta_i^\bT - (1-\delta) \bdelta_{i-1}^\bT A_{\min,i} \right) = 0.
    \end{align}
    Therefore,
    \begin{align}
        A_{\min,i} = \frac{1}{1-\delta}{\left(\bE \bdelta_{i-1}\bdelta_{i-1}^\bT\right)^{-1}  \bE  \bdelta_{i-1}\bdelta_{i}^\bT}
        \label{eq:Amin}
    \end{align}
    In turn, using~(\ref{eq:bdelta}), recursion~(\ref{eq:recursion_adaptive}) can be modified into
    \begin{align}
        \bdelta_i = (1-\delta)A_\star^\bT \bdelta_{i-1} + \delta\left(\bL_i - \frac 1M \sum_{j=i-M}^{i-1}\bL_{j}\right)
        \label{eq:bd_rec}
    \end{align}
    so that
    \begin{align}
        A_\star =&\; \frac{1}{1-\delta}\left(\bE \bdelta_{i-1}\bdelta_{i-1}^\bT\right)^{-1}\nonumber\\
        &\;\times \Bigg( \bE  \bdelta_{i-1}\bdelta_{i}^\bT - \delta \bE  \bdelta_{i-1}\Bigg(\bL_i - \frac 1M \sum_{j=i-M}^{i-1}\bL_{j}\Bigg)^\bT\Bigg)
        \label{eq:Astar}
    \end{align}
    Subtracting~(\ref{eq:Amin}) and~(\ref{eq:Astar}) gives:
    \begin{align}
        A_{\min,i} - A_\star =&\; \frac{\delta}{1-\delta}\left(\bE \bdelta_{i-1}\bdelta_{i-1}^\bT\right)^{-1} \nonumber\\ 
        &\; \times \bE \bdelta_{i-1} \left(\bL_i - \frac 1M \sum_{j=i-M}^{i-1}\bL_{j}\right)^\bT
        \label{eq:exp33}
    \end{align}
    Iterating~(\ref{eq:recursion_adaptive}), we get the following representation for $\blambda_i$:
    \begin{align}
        \blambda_i = (1-\delta)^M (A_\star^M)^\bT \blambda_{i-M} + \delta \sum_{t=0}^{M-1} (1-\delta)^t (A_\star^t)^\bT \bL_{i-t}.
        \label{eq:l_representation}
    \end{align}
    Since $\bL_i$ are i.i.d and using~(\ref{eq:exp33}) and the definition of $\bdelta_i$ in~(\ref{eq:bdelta}), we rewrite the expectation in~(\ref{eq:exp33}) as:
    \begin{align}
        &\bE \bdelta_{i-1}\left(\bL_i - \frac 1M \sum_{j=i-M}^{i-1}\bL_{j}\right)^\bT \nonumber\\
        &= \bE \left(\blambda_{i-1} - \frac 1M \sum_{j=i-M}^{i-1} \blambda_{j-1}\right)\left(\bL_i - \frac 1M \sum_{j=i-M}^{i-1}\bL_{j}\right)^\bT \nonumber\\
        &\overset{(\ref{eq:l_representation})}{=} \bE \Bigg[\Big((1-\delta)^M(A_\star^M)^\bT\nonumber\\
        &\;\;- \frac{(1-\delta)^{M-1}(A_\star^{M-1})^\bT + \dots + (1-\delta) A_\star^\bT + I}M \Big) \blambda_{i-M-1} \nonumber\\
        &+ \delta \Big((1-\delta)^{M-1}(A_\star^{M-1})^\bT\nonumber\\
        &\;\;- \frac{(1-\delta)^{M-2}(A_\star^{M-2})^\bT + \dots + I}M \Big) \bL_{i-M} \nonumber\\
        &+ \delta \Big((1-\delta)^{M-2}(A_\star^{M-2})^\bT\nonumber\\
        &\;\;- \frac{(1-\delta)^{M-3}(A_\star^{M-3})^\bT + \dots + I}M \Big) \bL_{i-M+1} \nonumber\\
        & + \dots \nonumber\\
        &+ \delta \left( (1-\delta) A_\star^\bT - \frac IM\right) \bL_{i-2} \nonumber\\
        &+ \delta \bL_{i-1}\Bigg] \times \Bigg[ \bL_{i} - \frac{\bL_{i-1} + \dots + \bL_{i-M}}M \Bigg]^\bT\nonumber\\
        &= -\frac{\delta}{M} \Big( \frac IM + \frac 2M (1-\delta)A_\star^\bT +\dots +\nonumber\\
        &+ \frac{M-1}M (1-\delta)^{M-2}(A_\star^{M-2})^\bT \nonumber\\
        &+ (1-\delta)^{M-1} (A_\star^\bT)^{M-1}\Big) \R_{\bL}\nonumber\\
        &=O\left(1 / M\right) I.
        \label{eq:Astar_additive}
    \end{align}
    with
    \begin{align}
        \R_{\bL} = \bE \left(\bL_i - \bar \cL\right)\left(\bL_i - \bar \cL\right)^\bT
    \end{align}
    and where we used the fact that 
    \begin{align}
        &\;\bE \left[\blambda_{i-M-1} \left( \bL_{i} - \frac{\bL_{i-1} + \dots + \bL_{i-M}}M \right)^\bT\right] \nonumber\\
        &\;= \bE \blambda_{i-M-1} \bE \left( \bL_{i} - \frac{\bL_{i-1} + \dots + \bL_{i-M}}M \right)^\bT \nonumber\\
        &\;= \bE \blambda_{i-M-1} \left(\bar\cL - \bar\cL \right)^\bT = 0.
        \label{eq:pppppp2}
    \end{align}
    Thus, combining~(\ref{eq:bdeltabdelta_fin}), (\ref{eq:exp33}) and (\ref{eq:Astar_additive}), we find the following relation between $A_\star$ and $A_{\min}$:
    \begin{align}
        &\|A_{\min, i} - A_\star\|_{\rm F}^2 \nonumber\\
        \leq&\; \frac {\delta^2}{(1-\delta)^2} \left\|O\left(1/ M\right)I\right\|_{\rm F}^2 \left\|\left(\bE \bdelta_{i-1}\bdelta_{i-1}^\bT\right)^{-1}\right\|_{\rm F}^2\nonumber\\
        \leq&\; \frac{|\mathcal N| \delta^2}{ \tau^2 \delta^4(1-\delta)^6} O\left(1/{M^2}\right) = O\left(1/ \delta^2 {M^2}\right)
    \end{align}
    
\section{Proof of Theorem~\ref{thm:conv}}\label{apx:conv}
    Using recursion~(\ref{eq:recursion_adaptive}) for  $\blambda_i$ and the definition of $\widehat\bL_{i-1}$ in~(\ref{eq:loglikelihood_estimator}), the SGD step~(\ref{eq:descent}) can be rewritten as follows:
    \begin{align}
        \bA_i 
        =\;& \bA_{i-1} + \mu(1-\delta)\bdelta_{i-1}\nonumber\\
        &\;\times \left(\bdelta_i^\bT - (1-\delta) \bdelta_{i-1}^\bT \bA_{i-1} \right)\nonumber\\
        \overset{(\ref{eq:bd_rec})}{=}\;& \bA_{i-1} + \mu(1-\delta)\bdelta_{i-1}\nonumber\\
        &\;\times \Big( (1-\delta) \bdelta_{i-1}^\bT \left(A_\star - \bA_{i-1}\right) \nonumber\\
        &\;\;\;\;\;+ \delta \Big(\bL_i - \frac 1M \sum_{j=i-M}^{i-1} \bL_j \Big)^\bT\Big)\nonumber\\
        =\;& \bA_{i-1} + \mu(1-\delta)^2\bdelta_{i-1} \bdelta_{i-1}^\bT \left(A_\star - \bA_{i-1}\right) \nonumber\\
        &\; + \mu \delta (1-\delta) \bdelta_{i-1} \Big(\bL_i - \frac 1M \sum_{j=i-M}^{i-1} \bL_j \Big)^\bT
        \label{eq:a_rec}
    \end{align}
    Recall the deviation from the true combination matrix:
    \begin{align}
        \widetilde \bA_i = A_\star - \bA_i.
        \label{eq:a_dev}
    \end{align}
    Combining~(\ref{eq:a_rec}) with~(\ref{eq:a_dev}), we obtain:
    \begin{align}
        \widetilde \bA_i 
        =\;& \left(I - \mu(1-\delta)^2\bdelta_{i-1} \bdelta_{i-1}^\bT\right)\widetilde \bA_{i-1} \nonumber\\
        &\; - \mu \delta (1-\delta) \bdelta_{i-1} \Big(\bL_i - \frac 1M \sum_{j=i-M}^{i-1} \bL_j \Big)^\bT \nonumber\\
        =\;& \left(I - \mu(1-\delta)^2\bE \bdelta_{i-1} \bdelta_{i-1}^\bT\right)\widetilde \bA_{i-1} \nonumber\\
        &\; + \mu(1-\delta)^2\left(\bE \bdelta_{i-1} \bdelta_{i-1}^\bT - \bdelta_{i-1} \bdelta_{i-1}^\bT\right)\widetilde \bA_{i-1} \nonumber\\
        &\; - \mu \delta (1-\delta) \bdelta_{i-1} \Big(\bL_i - \frac 1M \sum_{j=i-M}^{i-1} \bL_j \Big)^\bT.
        \label{eq:a_rec2}
    \end{align}
    The mean-square deviation is then given by
    \begin{align}
       &\bE \|\widetilde \bA_i\|_{\textrm F}^2\nonumber\\
       \leq&\; \rho^2 \left(I - \mu(1-\delta)^2 \bE \bdelta_{i-1}\bdelta_{i-1}^\bT\right) \bE \|\widetilde \bA_{i-1}\|_{\textrm F}^2\nonumber\\
       &\; + \mu^2 (1-\delta)^4 \bE \Big[ \Big\|\Big(\bE \bdelta_{i-1}\bdelta_{i-1}^\bT - \bdelta_{i-1}\bdelta_{i-1}^\bT\Big)\Big\|_{\textrm F}^2 \nonumber\\
       &\;\;\;\;\; \times \|\widetilde \bA_{i-1}\|_{\textrm F}^2\Big]\nonumber\\
       &\; + \mu^2\delta^2(1-\delta)^2\bE \Big\|\bdelta_{i-1} \Big(\bL_i - \frac 1M \sum_{j=i-M}^{i-1} \bL_j \Big)^\bT\Big\|^2_{\textrm{F}} \nonumber\\
       &\; - 2 \mu \delta(1-\delta) \bE \Big[ \textrm{Tr} \Big(\bdelta_{i-1} \Big(\bL_i - \frac 1M \sum_{j=i-M}^{i-1} \bL_j \Big)^\bT \nonumber\\
       &\;\;\;\;\; \times \widetilde \bA_{i-1} ^\bT  \left(I - \mu (1-\delta)^2\bdelta_{i-1}\bdelta_{i-1}^\bT\right)\Big)\Big] \nonumber\\
       &\; + 2\mu(1-\delta)^2 \bE \Big[\textrm{Tr}\Big( \left(\bE \bdelta_{i-1}\bdelta_{i-1}^\bT - \bdelta_{i-1}\bdelta_{i-1}^\bT\right) \widetilde \bA_{i-1}\nonumber\\
       &\;\;\;\;\; \times \widetilde \bA_{i-1}^\bT \left(I - \mu(1-\delta)^2 \bE \bdelta_{i-1}\bdelta_{i-1}^\bT  \right)\Big)\Big] \nonumber \\
       =&\; \rho^2 \left(I - \mu(1-\delta)^2 \bE \bdelta_{i-1}\bdelta_{i-1}^\bT\right) \bE \|\widetilde \bA_{i-1}\|_{\textrm F}^2\nonumber\\
       &\; + \mu^2 (1-\delta)^4 \bE \Big[ \Big\|\Big(\bE \bdelta_{i-1}\bdelta_{i-1}^\bT - \bdelta_{i-1}\bdelta_{i-1}^\bT\Big)\Big\|_{\textrm F}^2 \nonumber\\
       &\;\;\;\;\; \times \|\widetilde \bA_{i-1}\|_{\textrm F}^2\Big]\nonumber\\
       &\; + \mu^2\delta^2(1-\delta)^2\bE \Big\|\bdelta_{i-1} \Big(\bL_i - \frac 1M \sum_{j=i-M}^{i-1} \bL_j \Big)^\bT\Big\|^2_{\textrm{F}} \nonumber\\
       &\; + 2 \mu^2 \delta(1-\delta)^3 \bE \Big[ \textrm{Tr} \Big(\bdelta_{i-1} \Big(\bL_i - \frac 1M \sum_{j=i-M}^{i-1} \bL_j \Big)^\bT \nonumber\\
       &\;\;\;\;\; \times \widetilde \bA_{i-1} ^\bT  \bdelta_{i-1}\bdelta_{i-1}^\bT\Big)\Big] \nonumber\\
       &\; + 2\mu(1-\delta)^2 \bE \Big[\textrm{Tr}\Big( \left(\bE \bdelta_{i-1}\bdelta_{i-1}^\bT - \bdelta_{i-1}\bdelta_{i-1}^\bT\right) \widetilde \bA_{i-1}\nonumber\\
       &\;\;\;\;\; \times \widetilde \bA_{i-1}^\bT \left(I - \mu(1-\delta)^2 \bE \bdelta_{i-1}\bdelta_{i-1}^\bT  \right)\Big)\Big]\nonumber\\
       &\; - 2 \mu \delta(1-\delta) \bE \Big[ \textrm{Tr} \Big(\bdelta_{i-1} \Big(\bL_i - \frac 1M \sum_{j=i-M}^{i-1} \bL_j \Big)^\bT\widetilde \bA_{i-1} \Big)\Big]
       \label{eq:to_bound}
    \end{align}
    where $\rho(\cdot)$ denotes the spectral radius of its matrix argument. Considering the first term:
     \begin{align}
       &\;\rho^2 \left(I - \mu(1-\delta)^2 \bE \bdelta_{i-1}\bdelta_{i-1}^\bT\right)\nonumber\\
       =&\; \max \Big\{ \lambda_{\min}^2\left(I - \mu(1-\delta)^2 \bE \bdelta_{i-1} \bdelta_{i-1}^\bT \right), \nonumber\\
       &\;\;\;\;\;\;\;\;\;\;\; \lambda_{\max}^2\left(I - \mu(1-\delta)^2 \bE \bdelta_{i-1} \bdelta_{i-1}^\bT \right)\Big\} \nonumber\\
       =&\; \max \Big\{ \left(1 - \mu(1-\delta)^2 \lambda_{\min}\left(\bE \bdelta_{i-1} \bdelta_{i-1}^\bT \right)\right)^2, \nonumber\\
       &\;\;\;\;\;\;\;\;\;\;\;\; \left(1 - \mu(1-\delta)^2 \lambda_{\max}\left(\bE \bdelta_{i-1} \bdelta_{i-1}^\bT \right)\right)^2\Big\} \nonumber\\
       =&\; \max \left\{(1-\mu\nu)^2, (1-\mu\kappa)^2 \right\} \nonumber\\
       \leq&\; 1 - 2\mu\nu + \mu^2\kappa^2 = 1 - 2 \mu\nu + O(\mu^2),
       \label{eq:alpha_derivation}
    \end{align}
    where the last inequality holds since $\nu \leq \kappa$ by definitions~(\ref{eq:nu}) and (\ref{eq:kappa}). We omit the time indices in $\nu_i$ and $\kappa_i$ since we are assuming that $\blambda_i$ has reached the steady-state distribution~(\ref{eq:blambda_lim_}), and $\nu$ and $\kappa$ are defined as:
    \begin{align}
        &\nu \triangleq (1-\delta)^2\lambda_{\min} \left(\lim_{i\rightarrow\infty}\bE \bdelta_{i-1}\bdelta_{i-1}^\bT \right)\nonumber\\
        &\kappa \triangleq (1-\delta)^2\lambda_{\max} \left(\lim_{i\rightarrow\infty}\bE \bdelta_{i-1}\bdelta_{i-1}^\bT \right)
    \end{align}
    These constraints are positive by Lemma~\ref{lemma:risk} and exist due to bounded $\blambda_i$ in~(\ref{eq:lambda_bar_}). From \cite[Lemma 1]{shumovskaia2022explainability} it follows that:
    \begin{align}
        |\blambda_i| \preceq \bar \Lambda_i \triangleq&\; \left(1-\delta\right)^i \left(A_\star^\bT\right)^i|\Lambda_0| \nonumber\\
        &\;+ \delta b \sum_{t=0}^{i-1} \left(1-\delta\right)^{t}(A_\star^{t})^\bT \mathds{1}\mathds{1}^\bT
        \label{eq:lam_bound}
    \end{align}
    and the upper bound has a finite limit, $\lim_{i\rightarrow\infty} \bar \Lambda_i$. Therefore, we can derive another upper bound:
    \begin{align}
        |\bdelta_{i-1}\bdelta_{i-1}^\bT| \preceq &\;\frac 1 {M^2} \sum_{j_1=i-M}^{i-1} \left(\bar\Lambda_{i-1} + \bar\Lambda_{j_1}\right) \nonumber\\
        &\;\times \sum_{j_2=i-M}^{i-1} \left(\bar \Lambda_{i-1} + \bar \Lambda_{j_2}\right)^\bT \triangleq X.
        \label{eq:frakA}
    \end{align}
    Now, the expectation of the second term in~(\ref{eq:to_bound}) is bounded as follows:\footnotetext[5]{$(a-b)^2 \leq 2a^2 + 2b^2$.}
    \begin{align}
        &\;\mu^2 \bE \Big(\|(1-\delta)^2 \Big(\bE \bdelta_{i-1}\bdelta_{i-1}^\bT - \bdelta_{i-1}\bdelta_{i-1}^\bT\Big)\|_{\textrm F}^2 \nonumber\\
        &\;\times \|\widetilde \bA_{i-1}\|_{\textrm F}^2\Big)\nonumber\\
        \leq&^{\text{\footnotemark[5]}}\;2\mu^2 \bE \Big(\Big(\|(1-\delta)^2 \bE \bdelta_{i-1}\bdelta_{i-1}^\bT \|_{\textrm F}^2 \nonumber\\
        &\;\;\;\;\;\;+ \|(1-\delta)^2 \bdelta_{i-1}\bdelta_{i-1}^\bT \|_{\textrm F}^2 \Big)\nonumber\\
        &\; \times  \|\widetilde \bA_{i-1}\|_{\textrm F}^2\Big)\nonumber\\
        \leq&\;2\mu^2 \Big(\|(1-\delta)^2 \bE \bdelta_{i-1}\bdelta_{i-1}^\bT \|_{\textrm F}^2 + \|(1-\delta)^2 X \|_{\textrm F}^2 \Big)\nonumber\\
        &\; \times  \bE \|\widetilde \bA_{i-1}\|_{\textrm F}^2\nonumber\\
        \leq&\;4\mu^2 \|(1-\delta)^2 X \|_{\textrm F}^2\bE \|\widetilde \bA_{i-1}\|_{\textrm F}^2\nonumber\\
        =&\; O(\mu^2) \cdot \bE \|\widetilde \bA_{i-1}\|_{\textrm F}^2.
        \label{eq:trick1}
    \end{align}
    \noindent Now, we proceed with the third term in~(\ref{eq:to_bound}). Due to Assumption~\ref{asm:support}, we have bounded observations $\bL_i$ and $\blambda_i$ (see~(\ref{eq:lam_bound})), and we conclude that
    \begin{align}
        &\Big| \bdelta_{i-1}\Big( \bL_i - \frac 1M \sum_{j=i-M}^{i-1}\bL_j\Big)^\bT \Big| \nonumber\\
        &= \Big| \Big(\blambda_{i-1} - \frac 1M \sum_{j=i-M}^{i-1} \blambda_{j} \Big) \Big( \bL_i - \frac 1M \sum_{j=i-M}^{i-1}\bL_j\Big)^\bT \Big| \nonumber\\
        &\preceq \Big(|\blambda_{i-1}| + \frac 1M \sum_{j=i-M}^{i-1} |\blambda_{j}| \Big) \Big( |\bL_i| + \frac 1M \sum_{j=i-M}^{i-1}|\bL_j|\Big)^\bT  \nonumber\\
        &\preceq (\bar\Lambda + \bar \Lambda)(b \mathds 1 \mathds 1^\bT + b \mathds 1 \mathds 1^\bT)^\bT \nonumber\\
        & = O(1).
    \end{align}
    Thus, 
    \begin{align}
        &\;\mu^2\delta^2(1-\delta)^2\bE \Big\|\bdelta_{i-1} \Big(\bL_i - \frac 1M \sum_{j=i-M}^{i-1} \bL_j \Big)^\bT\Big\|^2_{\textrm{F}}\nonumber\\
        &\;\leq \mu^2 \delta^2 (1-\delta)^2 \cdot O(1) = \mu^2 \cdot O(\delta^2)
        \label{eq:trick2}
    \end{align}
    In turn, the fourth term becomes\footnotetext[6]{$2\textrm{ Tr}\left(AB^\bT\right) \leq \|A\|^2_{\textrm F} +\|B\|^2_{\textrm F}$.}:
    \begin{align}
        &\;2 \mu^2 \delta(1-\delta)^3 \bE \Big[ \textrm{Tr} \Big(\bdelta_{i-1} \Big(\bL_i - \frac 1M \sum_{j=i-M}^{i-1} \bL_j \Big)^\bT \nonumber\\
        &\; \times \widetilde \bA_{i-1} ^\bT  \bdelta_{i-1}\bdelta_{i-1}^\bT\Big)\Big]\nonumber\\
        \leq^{\text{\footnotemark[6]}} &\; \mu^2 \delta^2 (1-\delta)^6 \bE \Big\|\bL_i - \frac 1M \sum_{j=i-M}^{i-1} \bL_j \Big\|_{\textrm F}^2 \nonumber\\
        &\; + \mu^2 \bE \Big\| \bA_{i-1} ^\bT \bdelta_{i-1}\bdelta_{i-1}^\bT \bdelta_{i-1} \Big\|_{\textrm F}^2 \nonumber\\
        \overset{(a)}{=}&\; \mu^2 \delta^2(1-\delta)^6 \frac {M+1}{M} \Tr(\R_{\bL}) + O(\mu^2) \cdot \bE \|\widetilde \bA_{i-1} \|_{\textrm F}^2 \nonumber\\
        =&\; \mu^2 O(\delta^2) + O(\mu^2) \cdot \bE \|\widetilde \bA_{i-1} \|_{\textrm F}^2
        \label{eq:sim222}
    \end{align}
    where we use the same steps as in~(\ref{eq:trick1}), and $(a)$ follows from:
    \begin{align}
        &\;\bE \Big\| \frac 1M \sum_{j=i-M}^{i-1} \bL_j - \bL_i \Big\|_{\textrm F}^2 = \bE \Big\| \frac 1M \sum_{j=i-M}^{i-1} \left( \bL_j - \bL_i \right) \Big\|_{\textrm F}^2 \nonumber\\
        =&\; \bE \textrm{ Tr} \Bigg( \frac 1M \sum_{i=j-M}^{i-1} \left( \bL_j - \bL_i\right)\Bigg)\Bigg(\frac 1M \sum_{i=j-M}^{i-1} \left(\bL_j - \bL_i\right)\Bigg)^\bT \nonumber\\
        =&\;\frac 1{M^2} \sum_{j=i-M}^{i-1} \textrm{Tr } \bE \left(\left(\bL_j - \bL_i\right)\left(\bL_j - \bL_i\right)^\bT\right)\nonumber\\
        &\;\;\;\; + \frac 1{M^2} \sum_{j_1=i-M}^{i-1} \sum_{j_2 \neq j_1}^{i-1} \textrm{Tr } \bE \left(\left(\bL_{j_1} - \bL_i\right)\left(\bL_{j_2} - \bL_i\right)^\bT\right)\nonumber\\
        =&\;\frac 1{M^2} \sum_{j=i-M}^{i-1} \textrm{Tr } \left(\bE \bL_j \bL_j^\bT - 2 \bar\cL \bar\cL^\bT + \bE \bL_i \bL_i^\bT \right)\nonumber\\
        &\;\;\;\; + \frac 1{M^2} \sum_{j_1=i-M}^{i-1} \sum_{j_2 \neq j_1}^{i-1} \textrm{Tr } \left(\bE \bL_i \bL_i^\bT - \bar\cL \bar\cL^\bT \right)\nonumber\\
        =&\; \frac 2M \textrm{Tr}\left(\R_{\bL}\right) + \frac{M-1}{M} \Tr \left(\R_{\bL}\right) = \frac{M+1}{M} \Tr \left(\R_{\bL}\right)
    \end{align}
    To find the expectation of the second trace term in~(\ref{eq:to_bound}), consider first the conditional expectation, and then use Assumption~\ref{asm:independence}:
    \begin{align}
       &\;\bE \Big[\textrm{Tr}\Big( \left(\bE \bdelta_{i-1}\bdelta_{i-1}^\bT - \bdelta_{i-1}\bdelta_{i-1}^\bT\right) \widetilde \bA_{i-1}\widetilde \bA_{i-1}^\bT\nonumber\\
       &\;\;\;\;\; \times  \left(I - \mu(1-\delta)^2 \bE \bdelta_{i-1}\bdelta_{i-1}^\bT  \right)\Big)\Big|\bF_{i-M-1}\Big]\nonumber\\
       &\;=\textrm{Tr}\Big( \left(\bE \bdelta_{i-1}\bdelta_{i-1}^\bT - \bE\bdelta_{i-1}\bdelta_{i-1}^\bT\right)\nonumber\\
       &\;\;\;\;\; \times \bE \left[\widetilde \bA_{i-1} \widetilde \bA_{i-1}^\bT\big|\bF_{i-M-1}\right]\nonumber\\
       &\;\;\;\;\; \times \left(I - \mu(1-\delta)^2 \bE \bdelta_{i-1}\bdelta_{i-1}^\bT\right)\Big) \nonumber\\
       &\;= 0
    \end{align}
    where we use the fact that $\bdelta_{i-1}$ depends only on $\blambda_{i-M-1},\dots,\blambda_{i-1}$.
    Finally, consider the last term of~(\ref{eq:to_bound}). Using Assumption~\ref{asm:independence}, we obtain:
    \begin{align}
        &\;- 2 \mu \delta(1-\delta) \bE \Big[ \textrm{Tr} \Big(\bdelta_{i-1} \Big(\bL_i - \frac 1M \sum_{j=i-M}^{i-1} \bL_j \Big)^\bT\widetilde \bA_{i-1} \Big)\Big]\nonumber\\
        &\; = -2 \mu \delta(1-\delta) \nonumber\\
        &\;\;\;\;\;\;\; \times \bE \Big[ \textrm{Tr} \Big( \bE \Big(\bdelta_{i-1} \Big(\bL_i - \frac 1M \sum_{j=i-M}^{i-1} \bL_j \Big)^\bT \Big| \bF_{i-M-1} \Big) \nonumber\\
        &\;\;\;\;\;\;\;\;\;\;\;\; \times \bE \left(\widetilde \bA_{i-1} \big| \bF_{i-M-1} \right)\Big)\Big]\nonumber\\
        &\; \overset{(\ref{eq:pppppp2})}{=} -2 \mu \delta(1-\delta) \bE \Big[ \textrm{Tr} \Big( \bE \Big(\bdelta_{i-1} \Big(\bL_i - \frac 1M \sum_{j=i-M}^{i-1} \bL_j \Big)^\bT \Big) \nonumber\\
        &\;\;\;\;\;\;\; \times \bE \left(\widetilde \bA_{i-1} \big| \bF_{i-M-1} \right)\Big)\Big]\nonumber\\
        &\; \overset{(\ref{eq:Astar_additive})}{=} \mu \delta(1-\delta) \bE \Big[ \textrm{Tr} \Big( O(1/M)  \bE \left(\widetilde \bA_{i-1} \big| \bF_{i-M-1} \right)\Big)\Big]\nonumber\\
        &\; = \mu \delta(1-\delta) \bE \Big[ \textrm{Tr} \Big( O(1/M) \widetilde \bA_{i-1} \Big)\Big]\nonumber\\
        &\;\leq^{\text{\footnotemark[6]}} \mu \frac{(1-\delta)^2}{\delta} O(1/M^2) + \mu \delta^3 \bE \|\widetilde \bA_{i-1} \|_{\textrm F}^2 \nonumber\\
        &\; = \mu O(1/\delta M^2) + \mu \bE \|\widetilde \bA_{i-1} \|_{\textrm F}^2 \cdot O(\delta^3).
    \end{align}
    where in~(\ref{eq:Astar_additive}) the conditional expectation on $\bF_{i-M-1}$ is equal to the full expectation because (\ref{eq:pppppp2}) equals to zero for both full and conditional on $\bF_{i-M-1}$ expectations.
    Summarizing the derivations above, we can transform~(\ref{eq:to_bound}) into:
    \begin{align}
        \bE\|\widetilde{\bA}_i\|_{\textrm F}^2 \leq \alpha\bE\|\widetilde{\bA}_{i-1}\|_{\textrm F}^2 + \mu^2\gamma + \mu c,
        \label{eq:recursion_helper}
    \end{align}
    with $\gamma = O(\delta^2)$, $c = O\left( 1/\delta M^2\right)$ and $\alpha = 1 - \mu (2 \nu + O(\delta^3)) + O(\mu^2)$.
    Thus, 
    \begin{align}
        \bE\|\widetilde{\bA}_i\|_{\textrm F}^2 &\;\leq \alpha^i \|\widetilde{\bA}_0\|_{\textrm F}^2 + \left(\mu^2\gamma + \mu c\right)\sum_{t=0}^i \alpha^i \nonumber\\
        &\;= \alpha^i \|\widetilde{\bA}_0\|_{\textrm F}^2 + \left(\mu^2\gamma + \mu c\right) \frac {1-\alpha^i}{1-\alpha}.
    \end{align}
    For small enough $\mu$, $\alpha$ is strictly less than one. Hence, we obtain the following limiting MSD:
    \begin{align}
        &\;\limsup_{i\rightarrow\infty}\bE\|\widetilde{\bA}_i\|_{\textrm F}^2 \leq \frac {\mu^2\gamma + \mu c}{1-\alpha} \nonumber\\
        &\;= O(\mu) + O\left( 1 / \delta^3 M^2 \right)
    \end{align}
    
\section{Proof of Theorem~\ref{thm:conv2}}\label{apx:conv2}
    First, using recursion~(\ref{eq:recursion_adaptive}), we rewrite~(\ref{eq:loglikelihood_estimator}) as:
    \begin{align}
        \widehat \bL_{i-1}(\bA_i)
        &\;= \frac{1}{\delta M} \sum_{j=i-M}^{i-1}\left(\blambda_j - (1-\delta) \bA_i^\bT \blambda_{j-1}\right)\nonumber\\
        &\;= \frac 1{\delta M} \sum_{j=i-M}^{i-1} \left((1-\delta) \widetilde \bA_i ^\bT \blambda_{j-1} + \delta \bL_j\right)
    \end{align}
    Consider the mean square deviation of $\widehat \bL_{i}$:
    \begin{align}
        &\;\bE \|\widehat \bL_{i-1} - \bar \cL\|_{\textrm F}^2\nonumber\\
        = &\;\bE \Big\| \frac 1M \sum_{j=i-M}^{i-1} \bL_j - \bar \cL \Big\|_{\textrm F}^2 \nonumber\\
        &\; + \frac{(1-\delta)^2}{\delta^2} \bE \Big\|\widetilde \bA_i^\bT \cdot \frac 1M \sum_{j=i-M}^{i-1}  \blambda_{j-1}\Big\|_{\textrm F}^2 \nonumber\\
        &\; + 2\frac{1-\delta}\delta \bE \; \textrm{Tr} \Bigg( \frac 1M \widetilde \bA_i^\bT \sum_j \blambda_{j-1} \nonumber\\
        &\;\;\;\;\;\times \Bigg(\frac 1M \sum_j \bL_j - \bar\cL \Bigg)^\bT \Bigg).
        \label{eq:supp3}
    \end{align}
    Consider the first norm:
    \begin{align}
        &\;\bE \Big\| \frac 1M \sum_{j=i-M}^{i-1} \bL_j - \bar \cL \Big\|_{\textrm F}^2 = \bE \Big\| \frac 1M \sum_{j=i-M}^{i-1} \left( \bL_j - \bar \cL \right) \Big\|_{\textrm F}^2 \nonumber\\
        =&\; \bE \textrm{ Tr} \Bigg( \frac 1M \sum_{i=j-M}^{i-1} \left( \bL_j - \bar \cL\right)\Bigg)\Bigg(\frac 1M \sum_{i=j-M}^{i-1} \left(\bL_j - \bar \cL\right)\Bigg)^\bT \nonumber\\
        =&\;\frac 1{M^2} \sum_{i=j-M}^{i-1} \textrm{Tr } \bE \left(\left(\bL_j - \bar \cL\right)\left(\bL_j - \bar \cL\right)^\bT\right)\nonumber\\
        =&\; \frac 1M \textrm{Tr}\left(\R_{\bL}\right)
    \end{align}
    Let us study the second term of~(\ref{eq:supp3}) using Assumption~\ref{asm:independence} and Lemma~\ref{lemma:mean}. Conditioning on the history~(\ref{eq:history}), we get
    \begin{align}
        &\;\bE \left(\bE \left(\Big\|\widetilde \bA_i^\bT \cdot \frac 1M \sum_{j=i-M}^{i-1}  \blambda_{j-1}\Big\|_{\textrm F}^2\Big|\bF_{i-M-1}\right)\right)\nonumber\\
        \leq &\; \bE\left(\bE \left(\Big\|\widetilde \bA_i^\bT \Big\|_{\textrm F}^2 \cdot \Big\| \frac 1 M\sum_{j=i-M}^{i-1}  \blambda_{j-1}\Big\|_{\textrm F}^2\Big|\bF_{i-M-1}\right)\right) \nonumber\\
        = &\; \bE \Bigg(\bE \left(\Big\|\widetilde \bA_i^\bT\Big\|_{\textrm F}^2\Big|\bF_{i-M-1}\right) \nonumber\\
        &\;\;\;\;\;\times \bE \Big(\Big\| \frac 1 M \sum_{j=i-M}^{i-1}  \blambda_{j-1}\Big\|_{\textrm F}^2\Big|\bF_{i-M-1}\Big)\Bigg)
        \label{eq:pp2}
    \end{align}
    According to~(\ref{eq:av_lam}), we can bound:
    \begin{align}
        \bE \Big(\Big\| \frac 1 M \sum_{j=i-M}^{i-1}  \blambda_{j-1}\Big\|_{\textrm F}^2\Big|\bF_{i-M-1}\Big) \leq \|\bar\Lambda\|_{\textrm F}^2
    \end{align}
    Therefore,~(\ref{eq:pp2}) can be bounded as follows:
    \begin{align}
        &\bE \left(\bE \left(\Big\|\widetilde \bA_i^\bT \cdot \frac 1M \sum_{j=i-M}^{i-1}  \blambda_{j-1}\Big\|_{\textrm F}^2\Big|\bF_{i-M-1}\right)\right)\nonumber\\
        &\leq  \bE \| \widetilde \bA_i \|_{\textrm F}^2 \cdot O(1)
    \end{align}
    Now, consider the trace term in~(\ref{eq:supp3}): \begin{align}
        &\;\bE \textrm{ Tr} \left(\widetilde \bA_i^\bT \cdot   \frac 1M \sum_j \blambda_{j-1} \Big(  \frac 1M \sum_j \bL_j^\bT - \bar \cL^\bT \Big)\right) \nonumber\\
        =&\; 
        \bE\textrm{ Tr} \Bigg(\bE  \Bigg( \widetilde \bA_i^\bT \cdot \frac 1M \sum_j \blambda_{j-1} \nonumber\\
        &\;\times\Big(  \frac 1M \sum_j \bL_j^\bT - \bar\cL^\bT \Big)\Bigg| \bF_{i-M-1}\Bigg)\Bigg) \nonumber\\
        \overset{(a)}{=}&\; \bE \;\textrm{Tr} \Bigg( \bE \left( \widetilde \bA_i^\bT\big| \bF_{i-M-1}\right) \nonumber\\
        &\;\times \bE \Big( \frac 1M \sum_j \blambda_{j-1} \Big(\frac 1M \sum_j \bL_j^\bT - \bar\cL^\bT \Big)\Big| \bF_{i-M-1}\Big)\Bigg) \nonumber\\
        \overset{(b)}{=}&\; \bE \;\textrm{Tr} \left( \widetilde \bA_i^\bT \cdot O(1/M) I\right) \nonumber\\
        \leq^{\text{\footnotemark[6]}}&\; \frac 12 \bE \|\widetilde\bA_i\|_{\textrm F}^2 + O(1/M^2)
        \label{eq:supp4}
    \end{align}
    where $(a)$ holds due to Assumption~\ref{asm:independence} and $(b)$ holds due the trick similar to~(\ref{eq:Astar_additive}).
    Summarizing the derivations above, we can upper bound the expectation~(\ref{eq:supp3}) as:
    \begin{align}
        &\;\bE \|\widehat \bL_{i-1} - \bar \cL\|_{\textrm F}^2\nonumber\\
        \leq &\; \frac 1M \textrm{Tr} \left(\R_{\bL}\right) \nonumber\\
        &\;+  \frac{(1-\delta)^2}{\delta^2} \bE \| \widetilde \bA_i \|_{\textrm F}^2 \cdot O\left(1\right) \nonumber\\
        &\;+ \frac{1-\delta}{\delta} \left( \bE \| \widetilde \bA_i \|_{\textrm F}^2 + O(1/M^2)\right)\nonumber\\
        =&\; \frac 1M \textrm{Tr}\left(\R_{\bL}\right) + \frac{1}{\delta^2} \bE \| \widetilde \bA_i \|_{\textrm F}^2 \cdot O\left(1\right) + O(1/\delta M^2).
    \end{align}
    By Theorem~\ref{thm:conv}, we derive the following MSD in the limit:
    \begin{align}
        &\;\limsup_{i\rightarrow\infty} \bE \|\widehat \bL_{i-1} - \bar \cL\|_{\textrm F}^2 \nonumber\\
        &\;\leq \frac 1M \textrm{Tr} \left(\R_{\bL}\right) + O(\mu/\delta^2) + O\left(  1/\delta^5 M^2 \right)
    \end{align}

\section{}\label{apx:generator}
    In this section, we describe how we generate Bernoulli likelihood model parameters. Under hypothesis $\theta$, each agent $k$ receives observation $0$ with probability $p_k(\theta)$, and observation $1$ with probability $q_k(\theta) = 1-p_k(\theta)$. For hypotheses $\theta_0$, we fix the probabilities as follows:
    \begin{align}
        &p_k(\theta_0),\; q_k(\theta_0) = [0.3, 0.7]
    \end{align}
    And for all other $\theta_i\neq\theta_0$, we generate $p_k(\theta_i)$ and $q_k(\theta_i)$ with the following procedure:
    \begin{align}
        \varepsilon_{p_k},\; \varepsilon_{q_k} \sim \mathcal G(0, 1),
    \end{align}
    and
    \begin{align}
        p_k(\theta_i),\; q_k(\theta_i) =&\; \frac{1}{p_k(\theta_0) + \sigma_k^2\varepsilon_{p_k} + q_k(\theta_0) + \sigma_k^2\varepsilon_{q_k}}\nonumber\\
        &\;\times \left[p_k(\theta)+ \sigma_k^2\varepsilon_{p_k}, q_k(\theta_0) + \sigma_k^2\varepsilon_{q_k}\right] 
    \end{align}
    For ``non-influential" agents, we take $\sigma_k^2 = 0.05$, while for ``influential" ones we use $\sigma_k^2 = 0.5$ to enlarge the KL-divergences between different states. For additional comparison in Fig.~\ref{fig:influences}, we take $\sigma_k^2 = 0.2$ for ``influential" agents with smaller (but yet significant) KL-divergences.

\bibliographystyle{IEEEtran}
\bibliography{references}

\end{document}